\documentclass[aps,pra,twocolumn,reprint,amsmath,amssymb,graphicx,superscriptaddress]{revtex4-1}
\usepackage{graphicx,amsmath,amssymb,appendix,natbib,afterpage,amsfonts,hyperref,cleveref,latexsym,color,dcolumn,bm,mathtools}

\newcommand{\beq}{\begin{equation}}
\newcommand{\eeq}{\end{equation}}
\newcommand{\bea}{\begin{eqnarray}}
\newcommand{\eea}{\end{eqnarray}}

\providecommand{\moy}[1]{\langle #1 \rangle}
\providecommand{\bra}[1]{\langle #1 \rvert}
\providecommand{\ket}[1]{\lvert #1 \rangle}
\providecommand{\bbra}[1]{\langle\langle #1 \rvert\rvert}
\providecommand{\kket}[1]{\lvert\lvert #1 \rangle\rangle}
\providecommand{\braket}[2]{\langle #1 \rvert #2 \rangle}
\providecommand{\bbraket}[2]{\langle\langle #1 \rvert #2 \rangle\rangle}
\providecommand{\tr}[1]{\text{tr}\left[
 #1 \right]}
\providecommand{\parttr}[2]{\text{tr}_{#1}\left[
 #2 \right]}

\newcommand{\ketbra}[2]{\left| {#1} \right\rangle\left\langle {#2}\right|}
\newcommand{\ud}{\mathrm{d}}

\newcommand{\un}{\openone}

\providecommand{\dt}[1]{\frac{d#1}{dt}}
\providecommand{\commut}[2]{[#1,#2]}

\newcommand{\sA}{\mathcal{A}}
\newcommand{\sB}{\mathcal{B}}
\newcommand{\sC}{\mathcal{C}}
\newcommand{\sD}{\mathcal{D}}
\newcommand{\sP}{\mathcal{P}}
\newcommand{\sQ}{\mathcal{Q}}
\newcommand{\sV}{\mathcal{V}}

\newcommand{\sU}{\mathcal{U}}
\newcommand{\sL}{\mathcal{L}}
\newcommand{\sH}{\mathcal{H}}
\newcommand{\sG}{\mathcal{G}}
\newcommand{\sX}{\mathcal{X}}
\newcommand{\sbH}{\mathbfcal{H}}

\newcommand{\Leff}{\sL_{\text{eff}}}

\newcommand{\M}{\sU_{\mathrm{SWAP}}}
\newcommand{\ran}[1]{\text{ran}[#1]}

\graphicspath{ {/} }

\usepackage{tikz}
\usepgflibrary{arrows}
\usetikzlibrary{decorations}
\usetikzlibrary{decorations.pathmorphing,patterns}
\usetikzlibrary{scopes}
\usetikzlibrary{shapes, matrix}
\DeclareMathAlphabet\mathbfcal{OMS}{cmsy}{b}{n}

\begin{document}


\title{Projection based adiabatic elimination of bipartite open quantum  systems}


\author{Ibrahim Saideh}
\affiliation{Laboratoire Mat\'eriaux et Ph\'enom\`enes Quantiques,
Universit\'e Paris Diderot, CNRS UMR 7162, 75013, Paris, France.
Universit\'e Paris-Saclay, 91405 Orsay,
France}
\author{Daniel Finkelstein-Shapiro}
\affiliation{Division of Chemical
Physics, Lund University, Box 124,
221
00
Lund, Sweden}
\author{Camille No\^us}
\affiliation{Cogitamus Laboratory}
\email{arne.keller@u-psud.fr}
\author{T\~onu Pullerits}
\affiliation{Division of Chemical
Physics, Lund University, Box 124,
221
00
Lund, Sweden}

\author{Arne Keller}
\affiliation{Laboratoire Mat\'eriaux et Ph\'enom\`enes Quantiques,
Universit\'e Paris Diderot, CNRS UMR 7162, 75013, Paris, France.
Universit\'e Paris-Saclay, 91405 Orsay, France}
\email{arne.keller@u-psud.fr}

\begin{abstract}
Adiabatic elimination methods allow the reduction of the space dimension needed to describe systems dynamics which exhibits separation of time scale.
For open quantum system,  it  consists in  eliminating the fast part assuming it has almost instantaneously reached its steady-state and obtaining an approximation of  the evolution of the slow part. These methods can be applied to eliminate a linear subspace within the system Hilbert space, or alternatively to eliminate a fast subsystems in a bipartite quantum system. In this work, we extend an adiabatic elimination method used for removing fast degrees of freedom within a open quantum system (Phys. Rev. A \textbf{2020}, \textit{101}, 042102) to eliminate a subsystem from an open  bipartite quantum system. As an illustration, we apply our technique to a dispersively coupled two-qubit system  and in the case of the open Rabi model.
\end{abstract}

\maketitle

\section{Introduction}
Adiabatic elimination is a method whereby the fast degrees of freedom of a system are removed while retaining an effective description of the slow degrees of freedom. This simplification can be very useful to obtain tractable and intuitive equations when only a coarse-grained or long times description is desired \cite{Haken1975,Haken1977,Lax1967,Cohen-tannoudji1992, Paulisch_2014,Brion2007,
You2003,Nagy2010,Douglas2015}, depending on if the target system has a  conservative  \cite{Paulisch_2014,Sinatra1995,Brion2007} or a dissipative \cite{Azouit2016,Azouit2017,AzouitThesis2017,Azouit_structure-preserving_2017,Sarlette2020,Mirrahimi2009,Reiter2012,Kessler2012} evolution.
There are two classes of manifolds on which adiabatic elimination has been applied, i) those that consist of levels within a subsystem, for example the excited states of an atom, and ii) those that consist of a separate subsystem, such as ancillary qubits or measuring devices. For a slow and fast parts described by Hilbert spaces $\sH^{(A)}$ and $\sH^{(B)}$, the first case corresponds to the Hilbert space $\sH = \sH^{(A)}\oplus \sH^{(B)}$ (direct sum) while the second case corresponds to the Hilbert space $\sH = \sH^{(A)}\otimes \sH^{(B)}$ (tensor product).
Adiabatic elimination is useful in developing protocols for dissipative state preparation in ion traps \cite{Lin2013,Albert2019}, reservoir engineering \cite{Pastawski2011} and the description of measurement devices \cite{Cernotik2015}.
The simplicity of the resulting equations can also be computationally advantageous in the study of quantum phase transitions where the size of the system is cumbersomly large  \cite{Ciuti2018}.

There are several approaches to obtain effective operators, ranging from perturbative expansions of the Liouville operator \cite{Reiter2012,Kessler2012}, the corresponding Kraus maps \cite{Mirrahimi2009,Azouit2016,Azouit2017,
Azouit_structure-preserving_2017,AzouitThesis2017}, the resolvent \cite{Finkelstein-Shapiro2020}, or using stochastic methods \cite{Cernotik2015}.
Eliminating a fast subsystem (that forms a tensor product with the slow subsystem) is typically done with a partial trace over the fast subsystem. This can result in a set of hierarchical equations that allows error estimation and correcting the approximation as the slow and fast timescales get closer. Importantly, the expansion can be built to preserve the Lindblad structure and as a consequence the physicality of the map  \cite{Azouit2017,Lesanovsky2013,Marcuzzi2014}.
The procedure for eliminating sublevels within a subsystem (direct sum with the slow subsystem) is best carried out with Feshbach projections \cite{Reiter2012,Finkelstein-Shapiro2020}. However, as the fast-slow separation breaks down, or when incoherent pumping channels exist, the population of the fast subsystem becomes non-negligible (i.e. there can be a finite fraction of population in the excited states). When this happens, the exact time evolution of the slow part becomes non-trace preserving. The loss of trace can be corrected using contour integral methods \cite{Finkelstein-Shapiro2020}. It would be however advantageous to have a method that can handle both classes of fast manifolds. This is more important considering that systems from atomic physics are inspiring a number of chemical versions that have much more complicated Hamiltonians and it would be ideal to transform them into effective operators for a direct comparison to the atomic physics counterparts \cite{Castellini2018,Finkelstein-Shapiro2020_plex,Ribeiro2018}.

In this work, we extend the methodology developed in Ref.~\cite{Finkelstein-Shapiro2020} to
bipartite open quantum systems whose dynamics are described by a Lindblad
operator~\cite{Lindblad1976,Gorini1976}. 
We use the projection operator method suggested by Knezevic and Berry \cite{Knezevic2002} in order to derive equations for a slow subsystem $A$ coupled to a fast subsystem $B$. The paper is organized as follows.
We first recall the main results of Ref~\cite{Finkelstein-Shapiro2020}. We then apply it to the general bipartite case to obtain a recipe for describing the slow subsystem. Finally, we illustrate the method in the case of a spin dispersively coupled to a second highly dissipative driven spin and to describe the dynamics of the open Rabi model.
%

\section{Theory}
\subsection{Adiabatic elimination through projectors techniques}

Let $\rho(t)$ be the density operator on the Hilbert space $\sH$ describing
the quantum state of the system at time $t$. We suppose that
the evolution of $\rho(t)$ is
generated by a Lindblad operator
$\mathcal{L}$: $\dot{\rho}(t)
= \mathcal{L}\rho(t)$.
We define the Hilbert
space $\mathbfcal{H}$ of operator
$O$ on $\mathcal{H}$,  equipped
with  the scalar product
$\tr{O_1^\dagger O_2}$.
We first recall the main results presented in Ref~\cite{Finkelstein-Shapiro2020} related to projector techniques.  Let $\sP$ the  projector such that
$\rho_s(t) = \sP\rho(t)$
is the the slow part of the system and write $\sQ = \un - \sP$, where $\un$
is the identity operator on $\mathbfcal{H}$. Let $\sG(z) = (z-\sL)^{-1}$ be the resolvent of the Lindblad operator $\sL$. Operators like $\sP, \sQ, \sL$ or $\sG$ are operators on $\sbH$. They are sometimes called super-operators. They are here denoted with calligraphic letter,  to distinguish them from operators on $\sH$ (belonging to $\sbH$), like the density matrix $\rho$.

We define the effective Lindblad operator $\Leff(z)$, such that
$\sP \sG(z) \sP = (z-\Leff(z))^{-1}$. The effective Lindblad operator $\Leff(z)$ can be written as~:
\beq
\Leff(z) = \sP \sL \sP + \sP \sL \sQ \sG _0(z) \sQ \sL \sP,
\label{eq:Leff}
\eeq
where
\beq
\sQ \sG _0(z) \sQ =  (z-\sQ \sL \sQ)^{-1}.
\eeq
For any $\rho(t=0)$, such that
$\sQ\rho(t=0) = 0$, the slow dynamics inside $\sP\sbH$ can be obtained with the inverse Laplace transform as:
\beq
\rho_s(t) = \frac{1}{2\pi i}\int_{D} \ud z e^{zt} \sP G(z) \sP \rho_s(t=0),
\label{eq:invLaplaceTransf}
\eeq
where $\rho_s(t=0) = \sP\rho(t=0)$ and the integral
on the complex plane is performed on a straight line $D = \left\{z \in \mathbb{C}; \Re{z}=a>0\right\}$.
At this point no approximation has been made. Eq.~\eqref{eq:invLaplaceTransf} gives the exact dynamics inside $\sP\sbH$, as long as the initial condition is also inside $\sP\sbH$, that is $\sQ \rho(t=0)=0$. Because $\Leff(z)$ captures the effect of the dynamics in $\sQ\sbH$, it is a nonlinear superoperator, in the sense that $(z-\Leff(z))\vec{\nu}=0$ is a nonlinear eigenvalue problem.

The approximation of a slow dynamics of $\sP \rho(t)$,
with respect to the fast dynamics of $\sQ\rho(t)$ is equivalent to considering the dynamics inside $\sP \sbH$
in the vicinity of the stationary state reached in the limit $t\rightarrow \infty$.
In this long time limit only the $z \rightarrow 0$ limit will contribute  to the inverse Laplace transform
of Eq.~\eqref{eq:invLaplaceTransf}. We thus approximate $\Leff(z)$ to the lowest relevant order:
\beq
\Leff(z)\simeq \sL_0 + z\sL_1 + \cdots + z^n\sL_n
\label{eq:LeffAprox}
\eeq
where $\sL_0 =\Leff(z=0)$ and
$\sL_n = \frac{1}{n!}\left. \frac{\ud}{\ud z}\Leff(z) \right|_{z=0}$. Using the expression of
$\Leff(z)$ given by Eq.~\eqref{eq:Leff} allows to express $\sL_0$ and $\sL_n$ as:
\begin{align}
\sL_0 &=  \sP \sL \sP - \sP \sL\sQ(\sQ \sL \sQ)^{-1} \sQ \sL \sP \nonumber \\
\sL_n &= - \sP \sL \sQ\left(\sQ \sL \sQ\right)^{-(n+1)} \sQ \sL \sP.
\label{eq:L0L1}
\end{align}
In this work, we consider the approximation given by Eq.~\eqref{eq:LeffAprox} with $n\leq 1$ only, which is a standard approximation for most of the effective operators that are calculated explicitly. The systematic study of higher order approximations ($n>1$) will be considered in a future work.
Within the approximation given by Eq.~\eqref{eq:LeffAprox}, with $n=1$, the inverse Laplace transform of Eq.~\eqref{eq:invLaplaceTransf} can be computed explicitly. We obtain:
\beq
\rho_s(t) = \exp\left[(1 - \sL_1)^{-1} \sL_0 t \right] (1 - \sL_1)^{-1}\rho_s(t=0)
\label{eq:approxDynamics}
\eeq
The stationary state $\rho_f$ of this dynamics,  reached at $t\rightarrow +\infty$,  is in the kernel of $\sL_0$.
We note that although the dynamics described by Eq.~\eqref{eq:approxDynamics} is an  approximation,
the final reached stationary state $\rho_f$ is the exact one.

To conclude, after the adiabatic elimination of the fast part, the generator of the slow dynamics is approximatively given by
$(1 - \sL_1)^{-1}\sL_0$, $\dot{\rho}(t) = (1 - \sL_1)^{-1}\sL_0\rho(t)$,
where $\sL_0$ and $\sL_1$ can in principle be computed
using Eq.~\eqref{eq:L0L1}.
The hard part in these equations is the evaluation of the inverse
$(\sQ \sL \sQ)^{-1}$, which in the most general  case, as we will see later, can only be achieved through a perturbative expansion.

Theses results are very general, and require only the definition of a  projector $\sP$
and that the initial state fulfills the condition $\sQ\rho(t=0) = 0$. We note that $\sP$ doesn't have to be  hermitian, that is the projection does not need to be orthogonal.

In Ref.~\cite{Finkelstein-Shapiro2020},
this formalism was applied to the case where the slow and fast degrees of freedom correspond to a
 partition of the underlying Hilbert space in two complementary sub-spaces, that is
 $\sH = \sH^{(A)}\oplus \sH^{(B)}$. In the next section we will adapt this general formalism
 to the bipartite case where $\sH = \sH^{(A)}\otimes \sH^{(B)}$.

\subsection{Adiabatic elimination in a bipartite system }

We suppose that the state of the bipartite system at time $t$ is described by a density operator $\rho^{(AB)}(t)$ acting on
the Hilbert space $\sH = \sH^{(A)}\otimes \sH^{(B)}$. We consider that the dynamics of subsystem $A$ is very slow compared to the dynamics of subsystem $B$. We suppose that the exact stationary state in $\sbH$ is unique and that it
 is a product state $\rho_f = \rho_a\otimes\rho_b$, where $\rho_a \in \sbH^{(A)}$ and $\rho_b \in \sbH^{(B)}$ with $\sbH^{(i)}, i=A,B$, the Hilbert space of operators on $\sH^{(i)}$.
 As the the dynamics of $B$ subsystem is very fast, we suppose that it is a good approximation to consider that at $t=0$, $\rho^{(AB)}(t=0) = \rho^{(A)}_0\otimes\rho_b$. In other word, we consider that $B$ reaches its steady state instantaneously in the time scale of the subsystem $A$.   We thus define $\sP$ as
\beq
\sP \rho^{(AB)}(t) = \parttr{B}{\rho^{(AB)}(t)}\otimes \rho_b,
\eeq
where $\parttr{B}{}$ denotes the partial trace over $B$.
The reduced density operator $\rho^{(A)}(t)$ in $\sbH^{(A)}$ can then be obtained as
  $\rho^{(A)}(t) = \parttr{B}{\sP \rho^{(AB)}(t)}$.

For the purpose of simplifying some
expressions and calculations, it will
be
useful to use the operator-vector
isomorphism~\cite{Havel2003} which
maps
each element of $\mathbfcal{H}$
to a vector in
$\mathcal{H}\otimes\mathcal{H}$ as
follows.
An operators such as $\ket{a}\bra{b}
\in \mathbfcal{H}$ is mapped to
the vector
$\ket{\overline{b}}\otimes\ket{a}$ in
the
$\mathcal{H}\otimes\mathcal{H}$
Hilbert
space, where $\ket{\overline{b}}$ is
the complex conjugate of $\ket{b}$.
Consequently,  any
$n\times n$ density matrix $\rho \in
\mathbfcal{H}$ is mapped to a column
vector $\kket{\rho} \in
\mathcal{H}\otimes\mathcal{H}$,
with $n^2$
elements, by stacking the columns of
the $\rho$ matrix.
Under this isomorphism, super-operators on $\sbH$ are mapped to operators on
$\mathcal{H}\otimes\mathcal{H}$. In particular,
the super-operator $\mathcal{O}$ performing the operation
$\rho \rightarrow O_1\rho O_2^{\dagger}$, with $O_1$
and
$O_2$
operators in
$\mathbfcal{H}$, is mapped to
$\kket{\rho}\rightarrow \overline{O}_2\otimes O_1
\kket{\rho}$,
where $\overline{O}$ denotes the
complex
conjugate of $O$;
that
is $\overline{O} =
\left(O^{\dagger}\right)^T$, where
$O^{\dagger}$ is the
adjoint and $O^{T}$ is the transpose
of
$O$.
In this way, the scalar product
$\tr{\rho_1^\dagger\rho_2}$ between
two
operators $\rho_1$ and $\rho_2$ in
$\mathbfcal{H}$ is equal to the usual
scalar  product
$\bbraket{\rho_1}{\rho_2}$
in
$\mathcal{H}\otimes\mathcal{H}$. Some
useful remarks can be made.
The identity operator $\un$ in
$\mathbfcal{H}$ is mapped to
the
maximally entangled state
\beq
\kket{\un} =
\sum_k\ket{k}\otimes\ket{k}
\label{eq:mapIdentity}
\eeq
in
$\mathcal{H}\otimes\mathcal{H}$,
where
$\left\{\ket{k}\right\}$ is an
orthonormal basis of $\mathcal{H}$.
We also note that the usual density
matrix
normalization $\tr{\rho} = 1$ does
not
correspond to the normalization
induced by the scalar product
$\tr{\rho^2}=1$ (except in the case
of
a
pure state). Using the previous
remark,
we have that $\tr{\rho} =
\tr{\un \rho} = 1$ is mapped to
$\bbraket{\un}{\rho} = 1$.
For our bipartite case, $\mathcal{H} =
\mathcal{H}^{(A)}\otimes
\mathcal{H}^{(B)}$. Therefore,
an operator in $\mathbfcal{H}$ as
$\ket{a_1}\bra{a_2}\otimes
\ket{b_1}\bra{b_2}$, where $\ket{a_i
(b_i)} \in
\mathcal{H}^{(A)}(\mathcal{H}^{(B)})
(i=1,2)$, is mapped to
$\ket{\overline{a_2}}\otimes\ket{a_1}\otimes
\ket{\overline{b_2}}\otimes\ket{b_1}$.
The partial trace over $\mathcal{H}^{(B)}$,
$\parttr{B}{\rho} \in \mathbfcal{H}^{(A)}$ is mapped to
$\bbraket{\un^{(B)}}{\rho} \in \mathcal{H}^{(A)}\otimes \mathcal{H}^{(A)}$,
where $\ket{\un^{(B)}}=
\sum_k\ket{k}\otimes\ket{k}$ in
$\sH^{(B)}\otimes\sH^{(B)}$,
where
$\left\{\ket{k}\right\}$ is now  an
orthonormal basis of $\sH^{(B)}$.

Consequently, the operator $\sP \rho^{(AB)} \in \sbH$ is mapped to the vector
$\bbraket{\un^{(B)}}{\rho^{(AB)}}\otimes \kket{\rho_b} \in \sH\otimes\sH$.
The projector $\sP$ acting on  $\sH\otimes\sH$ can thus be written as:
\beq
\sP = \un^{(2A)}\otimes \sP^{(2B)} \text{ with } \sP^{(2B)} = \kket{\rho_b}\bbra{\un^{(B)}}
\label{eq:Pbipart}
\eeq
where $\un^{(2A)}$ is the identity operator on  $\sH^{(A)}\otimes\sH^{(A)}$.
The operator $\sQ = \un - \sP$ simply reads as:
\beq
\sQ = \un^{(2A)}\otimes  \sQ^{(2B)} \text{ with } \sQ^{(2B)} = \un^{(2B)} -\sP^{(2B)},
\eeq
where $\un^{(2B)}$ is the identity operator on  $\sH^{(B)}\otimes\sH^{(B)}$.

In general, the Lindblad operator of the system can be split in 3 terms as follows:
\beq
\sL = \sL_A \otimes \un^{(2B)} +   \un^{(2A)}\otimes \sL_B  + \sL_{AB}
\eeq
where $\sL_{A(B)}$ is a Lindblad operator acting on the $A(B)$ part only.
The decomposition of $\sL$ with the help of $\sP$ and $\sQ$ reads:
\begin{align}
\sP\sL\sP &= \sL_A\otimes \sP_B + \sP \sL_{AB}\sP \\
\sP\sL\sQ &= \sP\sL_{(AB)}\sQ \\
\label{eq:QLPfull}\sQ \sL \sP &= \un^{(2A)}\otimes \sQ_B \sL_B \sP_B +  \sQ\sL_{(AB)}\sP \\
\label{eq:QLQ} \sQ \sL \sQ &= \sL_A\otimes \sQ_B +\un^{(2A)}\otimes \sQ_B \sL_B \sQ_B + \sQ \sL_{AB}\sQ
\end{align}
Where we have used the fact that $\bbra{\un^{(B)}}\sL_B = 0$, as $\sL_B$ is a trace preserving operator. In some specific cases,  it can be a good approximation to take as $\rho_B$, a stationary state of $\sL_B$. In that case
$\sL_B\sP_B = 0$ and $\sQ \sL \sP$ in Eq.~\eqref{eq:QLPfull} can be simplified as
\beq
\sQ \sL \sP =   \sQ\sL_{(AB)}\sP
\label{eq:QLPsimple}
\eeq
For computing $\sL_0$ using Eq.~\eqref{eq:approxDynamics}, the main difficulty resides in the inversion of $\sQ \sL \sQ$. In general this inversion can not be done explicitly, but  a perturbative expansion can give a good approximation. In many cases $\sQ\sL\sQ$ can be divided into two terms,
$\sQ\sL\sQ = \sD + \sV$, where the computation of $\sD^{-1}$ is easy, and $\sV$ is small compared to $\sD$. In that case we can write
\beq
\label{eq:perturbative_inverse}
\left(\sQ \sL \sQ\right)^{-1} = \sD^{-1}\sum_{n=0}^{\infty} \left(\sV \sD^{-1}\right)^n
\eeq
Retaining the terms up to $n=0$ or $n=1$ may give a good approximation.

In some cases, the term $\un^{(2A)}\otimes \sQ_B \sL_B \sQ_B$ in Eq.~\eqref{eq:QLQ} is dominant over the two other terms. In that case, approximating the inverse of  $\sQ\sL\sQ$ by
\beq
\left(\sQ \sL \sQ\right)^{-1}\simeq \un^{(2A)}\otimes \left(\sQ_B \sL_B \sQ_B\right)^{-1}
\eeq
can be sufficient as we will see in the examples in the next section. So, $\sL_0$ can be approximated by the following expression:
\begin{align}
\sL_0 &= \sL_A\otimes \sP_B + \sP \sL_{AB}\sP  + \nonumber\\
& \sP\sL_{(AB)}\sQ
\left[\un^{(2A)}\otimes \left(\sQ_B \sL_B \sQ_B\right)^{-1}\right] \times \nonumber\\
&\left[ \un^{(2A)}\otimes \sQ_B \sL_B \sP_B +  \sQ\sL_{(AB)}\sP \right]
\end{align}
This is the main result of this work.

\section{Examples}
\label{example}
We apply the formalism of the preceding section to two examples. We first address  the case of a strongly dissipative driven qubit $B$ dispersively coupled to a target qubit $A$. Then, as a second example, we consider the open Rabi model in the regime where the dynamics of the spin is very fast compared to the boson frequency.

\subsection{A 2-qubit system}
\label{sec:twoqubit}
This two qubits system has been considered previously by Azouit et in Ref.~\cite{Azouit2017} to test  another method of bipartite adiabatic elimination (note that in their work, it is  the $A$ spin which is the strongly dissipative spin).
It consists in a strongly dissipative driven qubit $B$ dispersively coupled to a target qubit $A$. This model is used in Ref.~\cite{Sarlette2020} to describe the continuous measurement of a harmonic oscillator excitation number (corresponding to system $A$)  by a spin (corresponding system $B$).
The Lindblad equation for the bipartite system can be written as:
\begin{equation}
\begin{split}
\dt{\rho}&=u\commut{\sigma_+^B-\sigma_-^B}{\rho}+\gamma \left(\sigma_-^B\rho\sigma_+^B-\dfrac{\sigma_+^B\sigma_-^B\rho+\rho\sigma_+^B\sigma_-^B}{2}\right)\\
&-i\chi\commut{\sigma_z^A\otimes\sigma_z^B}{\rho}
\end{split}
\label{eq:twoqubitsModel}
\end{equation}
where $\sigma_+ = \ket{1}\bra{0}$, $\sigma_- = \sigma_+^{\dagger}$ and $\ket{0},\ket{1}$ are the eignvectors of the Pauli matrix $\sigma_z$ with eigenvalues $-1,1$ respectively.
Defining the new parameters $\tau=ut$, $\chi'=\dfrac{\chi}{u}$, and $\gamma'=\dfrac{\gamma}{u}$ and using the column-vector isomorphism, we write the superoperator form of the Liouvillian as:
\begin{equation}
\label{eq:2qubitsLindbladian}
\begin{split}
\sL=&-i\left[\chi'\left(\openone^{A}\otimes\sigma_z^A\otimes \openone^B\otimes \sigma_z^B-{\sigma}_z^A\otimes\openone^{A}\otimes  \sigma_z^B\otimes\openone^B\right) \right. \nonumber\\
& -\left. \left(\openone^{2A}\otimes \openone^B\otimes \sigma_y^B+\openone^{2A}\otimes  \sigma_y^B\otimes\openone^B\right)\right]\nonumber\\
&+\gamma'\openone^{2A}\otimes \sigma_-^B\otimes\sigma_-^B \\
& -\dfrac{\gamma'}{2}\left[\openone^{2A}\otimes \sigma_+^B\sigma_-^B\otimes\openone^B+\openone^{2A}\otimes \openone^B\otimes \sigma_+^B\sigma_-^B\right] \nonumber
\end{split}
\end{equation}
where we have used the relations
\begin{equation}
\bar{\sigma}_y=-\sigma_y\,,\,\bar{\sigma}_+=\sigma_+\,,\,\bar{\sigma}_-=\sigma_-
\end{equation}

From now on, we will drop the prime in all the parameters $\gamma'$, $\chi'$.
As the qubit $A$ only comes into play through the Hamiltonian term $-i\chi\commut{\sigma_z^A\otimes\sigma_z^B}{\rho}$ (see Eq.~\eqref{eq:twoqubitsModel}), the kernel of the Lindblad operator is two dimensional, according to the two eigenvectors $\sigma_z^A$. Hence the kernel can be considered as the span of $\{\rho_{s0}^{(A)}\otimes\rho_{s0}^{(B)}, \rho_{s1}^{(A)}\otimes\rho_{s1}^{(B)}\}$,
where $\rho_{s0}^{(A)} = \ket{0}\bra{0}$, $\rho_{s1}^{(A)} = \ket{1}\bra{1}$ and (see Appendix for details):
\begin{align}
\rho_{s_1}^{(B)}&=\dfrac{\un}{2}+\frac{2 \gamma}{16 \chi^2 + \gamma^2 + 8}\sigma_x^{(B)}+ \frac{8 \chi}{16 \chi^2 + \gamma^2 + 8}\sigma_y^{(B)} \nonumber \\
&  - \frac{16 \chi^2 + \gamma^2}{32 \chi^2 + 2 \gamma^2 + 16}\sigma_z^{(B)} \\
\rho_{s_0}^{(B)}&=\dfrac{\un}{2}+\frac{2 \gamma}{16 \chi^2 + \gamma^2 + 8}\sigma_x^{(B)}- \frac{8 \chi}{16 \chi^2 + \gamma^2 + 8}\sigma_y^{(B)} \nonumber \\
& - \frac{16 \chi^2 + \gamma^2}{32 \chi^2 + 2 \gamma^2 + 16}\sigma_z^{(B)}
\end{align}

In order to avoid any unnecessary complications, we will assume that the qubit $A$ has an extremely slow dissipation rate that we will omit from our calculations, but will ensure the uniqueness of the steady state.  In other words, the steady state of the considered system will be $\rho_{ss}=\rho_{s_0}^{(A)}\otimes\rho_{s_0}^{(B)}$. We thus assume that the initial state is of the form $\rho_0=\rho^{(A)}_0\otimes \rho_{s_0}^{(B)}$, and we define the projector $\sP$ (see Eq.~\eqref{eq:Pbipart}):
\begin{equation}
\sP=\openone^{2A}\otimes \kket{{\rho_{s_0}^{(B)}}}\bbra{\un^{(B)}},
\end{equation}
where according to Eq.~\eqref{eq:mapIdentity},  $\kket{\un^{(B)}} =\ket{0}\otimes\ket{0}+\ket{1}\otimes\ket{1}$.
This ensures that $\sQ=\un-\sP$ verifies $\sQ \ket{\rho_0}=0$ as it should be.
In Appendix \ref{App2}, we calculate in detail all the quantities necessary to compute $\mathcal{L}_0$ and $\mathcal{L}_1$ as given by Eq.~\eqref{eq:L0L1} :
\begin{align}
\sP\sL\sP=&\ketbra{0,1}{0,1}\otimes\sA_{0,1}+\ketbra{1,0}{1,0}\otimes\sA_{1,0}\label{eq:2qubit_PLP}\\
\sP\sL\sQ=&\ketbra{0,1}{0,1}\otimes\sC_{0,1}+\ketbra{1,0}{1,0}\otimes\sC_{1,0}\label{eq:2qubit_PLQ}\\
\sQ\sL\sP=&\ketbra{0,1}{0,1}\otimes\sD_{0,1}+\ketbra{1,0}{1,0}\otimes\sD_{1,0}\nonumber\\
+&\ketbra{0,0}{0,0}\otimes\sD_{0,0}+\ketbra{1,1}{1,1}\otimes\sD_{1,1}\label{eq:2qubit_QLP}\\
\sQ\sL\sQ=&\ketbra{0,1}{0,1}\otimes\sB_{0,1}+\ketbra{1,0}{1,0}\otimes\sB_{1,0}\nonumber\\
+&\ketbra{0,0}{0,0}\otimes\sB_{0,0}+\ketbra{1,1}{1,1}\otimes\sB_{1,1}\label{eq:2qubit_QLQ}
\end{align}
where $\sX_{i,j}$, $\sX\in\lbrace \sA,\sB,\sC,\sD\rbrace,\;\text{and}\; i,j \in {0,1}$, are operators acting on $\mathbfcal{H}^{(B)}$ and we have simplified the notation as
$\ket{i,j} = \ket{i}\otimes\ket{j}   \in \mathcal{H}^{(A)}\otimes\mathcal{H}^{(A)} \quad (i,j = 0,1)$,
see Appendix.~\ref{App2}. From the block diagonal form of Eq.~\eqref{eq:2qubit_QLQ}, it is relatively easy to invert $\sQ \sL\sQ$ exactly. In addition, we are only interested in quantities of the form $\sP \sL\sQ (\sQ \sL\sQ)^{-n} \sQ \sL\sP$. Hence, from the form of $\sP \sL\sQ$ in Eqs.\eqref{eq:2qubit_PLQ}, we only need to calculate $\mathcal{B}_{0,1}^{-1}$ and $\mathcal{B}_{1,0}^{-1}$.

Using Eqs.~\eqref{eq:2qubit_PLP} to \eqref{eq:2qubit_QLQ} in Eq.~\eqref{eq:L0L1}, we can write $\sL_0$ and $\sL_1$ as:
\begin{equation}
\begin{split}
\sL_0&=\sP\sL\sP-\sP\sL\sQ\left(\sQ\sL\sQ\right)^{-1}\sQ\sL\sP \\
&=\ketbra{0,1}{0,1}\otimes\left[\sA_{0,1}-\sC_{0,1}\sB^{-1}_{0,1}\sD_{0,1}\right]\\
&+\ketbra{1,0}{1,0}\otimes\left[\sA_{1,0}-\sC_{1,0}\sB^{-1}_{1,0}\sD_{1,0}\right]
\end{split}
\end{equation}
\begin{equation}
\label{eq:L1_2qubit_final}
\begin{split}
\sL_1&=-\sP\sL\sQ\left(\sQ\sL\sQ\right)^{-2}\sQ\sL\sP \\
&=-\ketbra{0,1}{0,1}\otimes\left[\sC_{0,1}\sB^{-2}_{0,1}\sD_{0,1}\right]\\
&-\ketbra{1,0}{1,0}\otimes\left[\sC_{1,0}\sB^{-2}_{1,0}\sD_{1,0}\right]
\end{split}
\end{equation}
Since $\sP^{(B)}$ is a projector of rank $1$, we can write:
\begin{equation}
\begin{split}
\sA_{0,1}-\sC_{0,1}\sB^{-1}_{0,1}\sD_{0,1}&=\alpha_{0,1}\sP^{(B)}\\
\sA_{1,0}-\sC_{1,0}\sB^{-1}_{1,0}\sD_{1,0}&=\alpha_{1,0}\sP^{(B)}\\
\sC_{0,1}\sB^{-2}_{0,1}\sD_{0,1}&=\beta_{0,1}\sP^{(B)}\\
\sC_{1,0}\sB^{-2}_{1,0}\sD_{1,0}&=\beta_{1,0}\sP^{(B)}
\end{split}
\end{equation}
where (see Appendix.~\ref{App2}):
\begin{equation}
\begin{split}
\alpha_{0,1}&\equiv \alpha=\bbra{\un^{(B)}}\left(\sA_{0,1}-\sC_{0,1}\sB^{-1}_{0,1}\sD_{0,1}\right)\kket{\un^{(B)}}\\
\beta_{0,1}&\equiv\beta=\bbra{\un^{(B)}}\left(\sC_{0,1}\sB^{-2}_{0,1}\sD_{0,1}\right)\kket{\un^{(B)}}\\
\alpha_{1,0}&=\overline{\alpha}\,,\,\beta_{1,0}=\overline{\beta}
\end{split}
\end{equation}
and, see Appendix~\ref{App2}:
\begin{equation}
\alpha=-\zeta+i\xi.
\end{equation}

\begin{figure}
\includegraphics[width=0.5\textwidth]{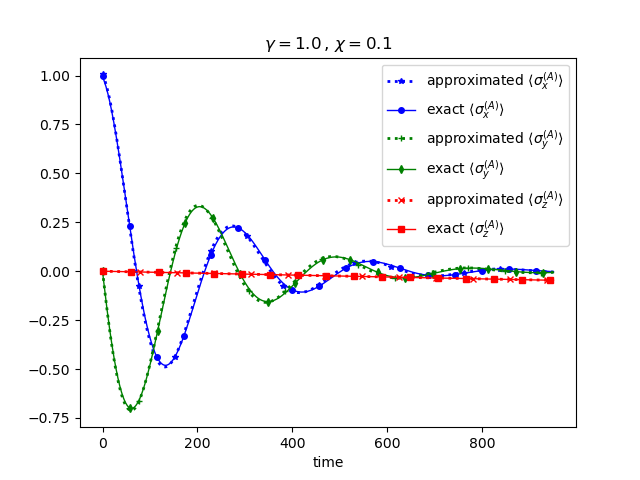}
\includegraphics[width=0.5\textwidth]{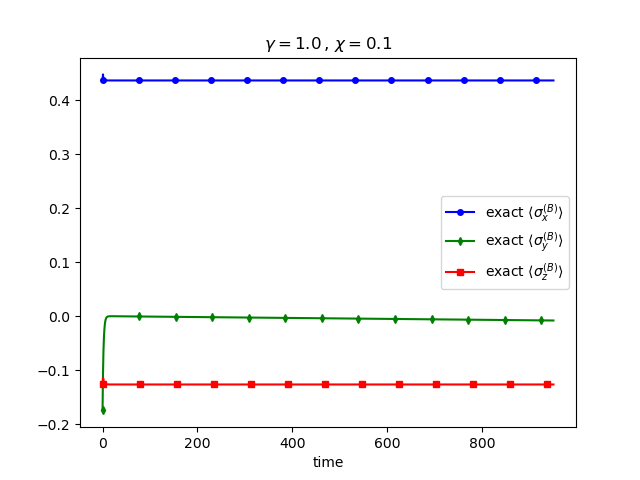}
\caption{Top: Evolution of the expectations values of the Pauli matrices $\moy{\sigma_x^A}$ (initial value = 1), $\moy{\sigma_y^A}$ (initial value=0) and $\moy{\sigma_z^A}$ (always zero) for  the slow spin $A$. Dashed line: Adiabatic elimination, continuous line: exact. Bottom: Evolution of the fast spin B. The initial state has been taken as $\ket{\phi_+}=\frac{1}{\sqrt{2}}\left(\ket{0}+\ket{1}\right)$.
}
\label{fig:example}
\end{figure}

With these variables and truncating~\eqref{eq:LeffAprox} to the zeroth order, the effective Liouville equation can be written in operator space as:
\begin{equation}
\dfrac{d}{d\tau}\rho^A=i\dfrac{\xi}{2}\left[\sigma_z^A,\rho^A\right]+\dfrac{\zeta}{2}\left(\sigma_z^A\rho^A\sigma_z^A-\rho^A\right)
\label{eq:approxL02qubit}
\end{equation}
Note that approximating the expression of $\xi$ and $\zeta$ (given in Appendix.~\ref{App2}) by their lowest order in  $\chi$ gives for Eq.~\eqref{eq:approxL02qubit} the same result as the one obtained by Azouit et al.~\cite{Azouit2017} using a completely different method.
Finally, it is straightforward to calculate $\left(\un-\sL_1\right)^{-1}$ exactly , given that $\sL_1$ (Eq.~\eqref{eq:L1_2qubit_final}) is block diagonal:
\begin{equation}
\begin{split}
\left(\un-\sL_1\right)^{-1}&=\ketbra{0,1}{0,1}\otimes\left[\frac{1}{1+\beta}\sP_B+\sQ_B\right]\\
&+\ketbra{1,0}{1,0}\otimes\left[\frac{1}{1+\overline{\beta}}\sP_B+\sQ_B\right]
\\
&+\left[\ketbra{0,0}{0,0}+\ketbra{1,1}{1,1}\right]\otimes \un^{(2B)}
\end{split}
\end{equation}
Let us define the modified initial state of the qubit $A$ as $\ket{\tilde{\rho}^{(A)}_0} = (1 - \sL_1)^{-1} {\rho}^{(A)}_0$:
\begin{equation}
\ket{\tilde{\rho}^{(A)}_0}=\rho_{0,0}\ket{0,0}+\rho_{1,1}\ket{1,1}+\frac{\rho_{0,1}}{1+\beta}\ket{0,1}+\frac{\rho_{1,0}}{1+\overline{\beta}}\ket{1,0},
\end{equation}
where $\rho_{0,0}, \rho_{1,1}$ represent population in the state $\rho^{(A)}_0$ while $\rho_{0,1}, \rho_{1,0}$ represent initial coherences. The physical meaning of this redefined initial density matrix is the state of qubit $A$ immediately after qubit $B$ has reached its steady-state. In this case this corresponds to a rescaling of the coherences. Using $\sQ_B\rho^{(B)}_{s_0}=0$, we can rewrite Eq.~\eqref{eq:approxDynamics} to describe the dynamics of the slow qubit as:
\begin{equation}
\ket{\rho^A(t)}=e^{\tilde{\sL}_0t}\ket{\tilde{\rho}^A_0}
\end{equation}
where $\tilde{\sL}_0=(1 - \sL_1)^{-1}\sL_0=\frac{\alpha}{1+\beta}\ketbra{0,1}{0,1}+\frac{\overline{\alpha}}{1+\overline{\beta}}\ketbra{1,0}{1,0}$. The evolution operator $U(t)=e^{\tilde{L}_0t}$ is simple to calculate:
\begin{equation}
\begin{split}
U(t)&=e^{-\zeta' t+i\xi' t}\ketbra{0,1}{0,1}+e^{-\zeta' t-i\xi' t}\ketbra{1,0}{1,0}\\
&+\ketbra{0,0}{0,0}+\ketbra{1,1}{1,1}
\end{split}
\end{equation}
where we have defined
\begin{equation}
\zeta'=-\Re{\frac{\alpha}{1+\beta}}\,,\,\xi'=\Im{\frac{\alpha}{1+\beta}}
\end{equation}
The evolution of the approximated expectation value of the Pauli matrices for qubit $A$ and $B$ for $\gamma=1$ and $\chi=0.1$ and with the initial state taken as $$\ket{\phi_+}=\frac{1}{\sqrt{2}}\left(\ket{0}+\ket{1}\right)$$ are compared in Figure \ref{fig:example} to the ones obtained through an exact full numerical propagation. We see that the adiabatic elimination captures the exact dynamics faithfully and that indeed qubit  $B$ reaches its steady-state before any appreciable dynamics in $A$ has taken place.\\

It is worth mentioning that when adiabatic elimination is valid, i.e. $\chi\ll \gamma$, the exact final state of the fast qubit $B$ is very close to the steady state of $\sL_{B}$:
\begin{equation}
\rho_{ss}^{(B)}=\lim_{\chi \to 0}\rho_{s_0}^{(B)}.
\end{equation}
Defining the projector $\sP_B=\kket{\rho_{ss}^{(B)}}\bbra{\un^{(2B)}}$ leads to considerable simplifications in Eq.~\eqref{eq:QLPfull} where the first term becomes zero. Thus, taking the interaction $\sQ\sL_{AB}\sQ$ to be small, we only need the term $n=0$ in Eq.~\eqref{eq:perturbative_inverse} and taking the zero order only in~\eqref{eq:LeffAprox}, one can check that it leads to the same Lindblad operator derived in~\cite{Azouit2017} which is enough to obtain a very good approximation in the adiabatic limit.

\subsection{Open Rabi model}
The open Rabi model has been considered recently by Garbe et al.~\cite{Garbe2020} in a quantum metrology context.
It consist in a spin-$\frac{1}{2}$ (with frequency $\Omega$) interacting with one bosonic mode (frequency $\omega_0$) of a cavity described by the following Hamiltonian:
\begin{equation}
H_R = \Omega\sigma_z + \omega_0a^\dagger a + \lambda(a+a^\dagger)\otimes\sigma_x,
\end{equation}
where $a$ ($a^\dagger$) is the annihilation (creation) operator of the bosonic mode.
The dynamics of the open Rabi model where the relaxation of the spin (at a rate $\Gamma'$) and the photon losses from the cavity (at a rate $\kappa'$) are taken into account is generated by the following Lindblad operator:
\begin{equation}
\label{eq:SB_Lindbaladian}
\sL\left(\rho\right)=-i[H_R,\rho]+\Gamma'\sD_{\sigma_-}\left(\rho\right)+\kappa' \sD_a\left(\rho\right)
\end{equation}
where we have used the notation
\[\sD_X\left(\rho\right)=X\rho X^\dagger-\frac{1}{2}X^\dagger X\rho-\frac{1}{2}\rho X^\dagger X.\]
We also assume that $\Gamma'\thicksim \Omega$ and $\kappa'\thicksim \omega_0$.
If we rescale the time by dividing the above equation by $\sqrt{\Omega\omega_0}$, and define:
\begin{equation}
g=\dfrac{\lambda}{\sqrt{\Omega\omega_0}}\,,\, \eta=\sqrt{\dfrac{\omega_0}{\Omega}} \,,\, \kappa=\dfrac{\kappa'}{\sqrt{\Omega\omega_0}}\,,\,\Gamma=\dfrac{\Gamma'}{\sqrt{\Omega\omega_0}}
\label{eq:rescale}
\end{equation}
we rewrite the Lindbladian:
\begin{equation}
\sL\left(\rho\right)=\sL_A\left(\rho\right)+\sL_{AB}\left(\rho\right)+\sL_B\left(\rho\right)
\end{equation}
with :
\begin{equation}
\begin{split}
\sL_A\left(\rho\right)&=-i\eta \commut{ a^\dagger a}{\rho}+\kappa \sD_a\left(\rho\right)\\
\sL_B\left(\rho\right)&=-i\dfrac{1}{\eta} \commut{ \sigma_z}{\rho}+\Gamma \sD_{\sigma_-}\left(\rho\right)\\
\sL_{AB}\left(\rho\right)&=-ig \commut{\left(a+a^\dagger\right)\otimes\sigma_x}{\rho}.
\end{split}
\end{equation}
It has been shown that, in the limit where $\omega_0\ll \Omega$, this model exhibits a quantum phase transition when $g$ increases~\cite{Hwang2015,Puebla2017,Hwang2018,Garbe2020}. The critical point corresponding to $g=1$ separates a normal phase ($g<1$) from a superadiant phase ($g>1$).  Here we show that our method can be used to obtain an effective Lindblad operator for the boson in the normal phase after the elimination of the fast spin. After rescaling,  the adiabatic limit $\omega_0\ll \Omega$ corresponds to $\eta\rightarrow 0$.

In the normal phase ($g<1$), the steady state of the system, which is the  kernel of the Lindblad operator given by Eq.~\eqref{eq:SB_Lindbaladian}, is separable and unique~\cite{Garbe2020}. It is straight forward to verify that the steady state of $\sL_B$ is:
 \begin{equation}
 \kket{\rho^{(B)}_{ss}}=\ket{0,0}
 \end{equation}
Following the same steps of the last example, see Appendix.~\ref{App.SB}, we define the projector $\sP_B=\kket{\rho^{(B)}_{ss}}\bbra{\un^{(2B)}}$ and we only calculate the term $\sQ\sL_B\sQ$, the inverse of which corresponds to $\sQ\sL\sQ^{-1}$ up to zeroth order in Eq.~\eqref{eq:perturbative_inverse}. Simple and straightforward calculations lead to the following Lindbladian evolution of the boson:
\begin{equation}
\begin{split}
\sL_0\left(\rho^{(A)}\right)=&-i\commut{\eta a^{\dagger}{a}-\dfrac{4g^2\eta}{\Gamma^2\eta^2+16}\left(a+a^{\dagger}\right)^2}{\rho^{(A)}}\\&+\kappa\sD_{a}\left(\rho^{(A)}\right)+\dfrac{4g^2\eta^2\Gamma}{\Gamma^2\eta^2+16}\sD_{\left(a+a^{\dagger}\right)}\left(\rho^{(A)}\right)
\end{split}
\end{equation}
which is exactly the formula derived in~\cite{Garbe2020} using a completely different method, where one should take into consideration that the parameters in~\eqref{eq:SB_Lindbaladian} are double those considered in~\cite{Garbe2020}.

\section{Conclusion}
We have derived a projection based adiabatic elimination method that works for bipartite systems. This work provides a direct connection to earlier work on adiabatic elimination of a subspace of the system Hilbert space~\cite{Finkelstein-Shapiro2020} so that in principle now subsystems as well as sublevels can be eliminated at the same time. We have illustrated this with two simple examples of two dispersively coupled spins and the open Rabi model. In both cases, using the lowest order approximations, we have obtain the same expressions that have been previously obtained by completely different methods. We expect that this work will find applications in the case of molecules in cavities where the cavity and part of the molecular levels could be adiabatically eliminated.
\appendix
\section{Detailed Calculation for the 2-qubit system}
\label{App2}
Here we present in detail all the calculations involved in the example presented in section~\ref{sec:twoqubit}: first we write $\kket{\rho_s^{(B)}}$ in the standard basis:
\begin{align}
\label{eq:App2_rhobs}
\kket{\rho_s^{(B)}}=&\dfrac{16 \chi^{2} + \gamma^{2} + 4}{16 \chi^{2} + \gamma^{2} + 8}\ket{0,0}+\dfrac{4}{16 \chi^{2} + \gamma^{2} + 8}\ket{1,1}\nonumber\\+&\dfrac{2\gamma+8i\chi}{16 \chi^{2} + \gamma^{2} + 8}\ket{0,1}+\dfrac{2\gamma-8i\chi}{16 \chi^{2} + \gamma^{2} + 8}\ket{1,0}
\end{align}
where in this appendix we use the notation $\ket{i,j} = \ket{i}\otimes\ket{j}$ to alleviate the complexity of mathematical expressions.
Then, we define the necessary projectors of the partial trace, namely:
\begin{equation}
\label{eq:App2_PB}
\sP^{(B)}=\kket{\rho_s^{(B)}}\bbra{\un^{(B)}}
\end{equation}
and
\begin{align}
\label{eq:App2_QB}
\sQ^{(B)}=&\un^{(2B)}-\sP^{(B)}\nonumber\\
&\frac{4}{16 \chi^{2} + \gamma^{2} + 8}\ketbra{\Psi_-}{\Psi_+}-\ketbra{\Psi_-}{1,1}\nonumber\\
&-\frac{8 i \chi + 2 \gamma}{16 \chi^{2} + \gamma^{2} + 8}\ket{0,1}\bra{\Psi_+}+\ketbra{0,1}{0,1}\nonumber
\\&+\frac{8 i \chi - 2 \gamma}{16 \chi^{2} + \gamma^{2} + 8}\ket{1,0}\bra{\Psi_+}+\ket{1,0}\bra{1,0}
\end{align}
where we have defined the following vectors:
\begin{align}
\ket{\Psi_+}&=\ket{0,0}+\ket{1,1}\nonumber\\
\ket{\Psi_-}&=\ket{0,0}-\ket{1,1}
\end{align}
Later on, it will be useful to define
\begin{align}
\ket{\Phi_+}&=\ket{0,1}+\ket{1,0}\nonumber\\
\ket{\Phi_-}&=\ket{0,1}-\ket{1,0}\nonumber\\
\ket{\Theta_+}&=\dfrac{2\gamma+8i\chi}{16 \chi^{2} + \gamma^{2} + 8}\ket{0,1}+\dfrac{2\gamma-8i\chi}{16 \chi^{2} + \gamma^{2} + 8}\ket{1,0}\nonumber\\
\ket{\Theta_-}&=\dfrac{2\gamma+8i\chi}{16 \chi^{2} + \gamma^{2} + 8}\ket{0,1}-\dfrac{2\gamma-8i\chi}{16 \chi^{2} + \gamma^{2} + 8}\ket{1,0}\nonumber\\
\ket{\Omega_+}&=\dfrac{16 \chi^{2} + \gamma^{2} + 4}{16 \chi^{2} + \gamma^{2} + 8}\ket{0,0}+\dfrac{4}{16 \chi^{2} + \gamma^{2} + 8}\ket{1,1}\nonumber\\
\ket{\Omega_-}&=\dfrac{16 \chi^{2} + \gamma^{2} + 4}{16 \chi^{2} + \gamma^{2} + 8}\ket{0,0}-\dfrac{4}{16 \chi^{2} + \gamma^{2} + 8}\ket{1,1}
\nonumber\\
\ket{\Omega^\perp_+}&=\dfrac{16 \chi^{2} + \gamma^{2} + 4}{16 \chi^{2} + \gamma^{2} + 8}\ket{1,1}-\dfrac{4}{16 \chi^{2} + \gamma^{2} + 8}\ket{0,0}\nonumber\\
\ket{\Omega^\perp_-}&=\dfrac{16 \chi^{2} + \gamma^{2} + 4}{16 \chi^{2} + \gamma^{2} + 8}\ket{1,1}+\dfrac{4}{16 \chi^{2} + \gamma^{2} + 8}\ket{0,0}
\end{align}
and the unitary matrix:
\begin{equation}
\M=\ketbra{0,0}{0,0}+\ketbra{1,1}{1,1}+\ketbra{1,0}{0,1}+\ketbra{0,1}{1,0}
\end{equation}
as well. From Eqs.~\eqref{eq:App2_rhobs},~\eqref{eq:2qubitsLindbladian} and~\eqref{eq:App2_QB}, we find that:
\begin{equation}
\label{eq:App2_PLPQ}
\begin{split}
\sP\sL\sP=&\ketbra{0,1}{0,1}\otimes\sA_{0,1}+\ketbra{1,0}{1,0}\otimes\sA_{1,0}\\
\sP\sL\sQ=&\ketbra{0,1}{0,1}\otimes\sC_{0,1}+\ketbra{1,0}{1,0}\otimes\sC_{1,0}\\
\sQ\sL\sP=&\ketbra{0,1}{0,1}\otimes\sD_{0,1}+\ketbra{1,0}{1,0}\otimes\sD_{1,0}\\
+&\ketbra{0,0}{0,0}\otimes\sD_{0,0}+\ketbra{1,1}{1,1}\otimes\sD_{1,1}\\
\sQ\sL\sQ=&\ketbra{0,1}{0,1}\otimes\sB_{0,1}+\ketbra{1,0}{1,0}\otimes\sB_{1,0}\\
+&\ketbra{0,0}{0,0}\otimes\sB_{0,0}+\ketbra{1,1}{1,1}\otimes\sB_{1,1}
\end{split}
\end{equation}
Where we have defined the following matrices on $\mathbfcal{H}^{(2B)}$
\begin{align}
\sA_{0,1}=& \frac{2 i \chi \left(16 \chi^{2} + \gamma^{2}\right)}{\left(16 \chi^{2} + \gamma^{2}+8\right)}\sP^{(B)}\,,\, \sA_{1,0}=\M\overline{\sA_{0,1}}\\
\sC_{0,1}=&-4 i \chi\ket{\rho_s^{(B)}}\bra{\Omega^\perp_+}\,,\, \sC_{1,0}=\M\overline{\sC_{0,1}}
\end{align}
\begin{widetext}
\begin{align}
\sD_{0,1}=&\frac{16 i \chi \left(16 \chi^{2} + \gamma^{2} + 4\right)}{\left(16 \chi^{2} + \gamma^{2} + 8\right)^{2}}\ketbra{\Psi_-}{\Psi_+}- \frac{4 i \chi \left(16 \chi^{2} + \gamma^{2}\right) }{\left(16 \chi^{2} + \gamma^{2} + 8\right)}\ketbra{\Theta_+}{\Psi_+}-2i\chi\ketbra{\Theta_-}{\Psi_+}\nonumber\\
\sD_{1,1}=&-4 i\chi \ketbra{\Theta_-}{\Psi_+}\,,\,\sD_{1,0}=\M\overline{\sD_{0,1}}\,,\, \sD_{0,0}=\mathbf{0}
\end{align}

\begin{align}
\label{App2:B01}
\sB_{0,1}=& \frac{2i\chi\left(16\chi^2+\gamma^2\right)}{\left(16\chi^2+\gamma^2+8\right)}\ketbra{\Psi_-}{\Omega^\perp_+}+
\gamma\ketbra{\Psi_-}{1,1}-\frac{\gamma}{2}\ketbra{0,1}{0,1}-\frac{\gamma}{2}\ketbra{1,0}{1,0}-\ketbra{\Psi_-}{\Phi_+}+\ketbra{\Phi_+}{\Psi_-}\nonumber\\
&+\frac{2i\chi\left(16\chi^2+\gamma^2\right)}{\left(16\chi^2+\gamma^2+8\right)}\ketbra{\Theta_-}{\Psi_-}+\frac{64i\chi\left(4i\chi-\gamma\right)}{\left(16\chi^2+\gamma^2+8\right)^2}\ketbra{1,0}{0,0}+\frac{16i\chi\left(4i\chi+\gamma\right)\left(16\chi^2+\gamma^2+4\right)}{\left(16\chi^2+\gamma^2+8\right)^2}\ketbra{0,1}{1,1}	\nonumber\\
\sB_{0,0}=&-\ketbra{\Psi_-}{\Phi_+}+\ketbra{\Phi_+}{\Psi_-}+\gamma\ketbra{\Psi_-}{1,1}+ \frac{4i\chi-\gamma}{2}\ketbra{0,1}{0,1}- \frac{ 4 i \chi+\gamma}{2}\ketbra{1,0}{1,0}\nonumber\\
\sB_{1,1}=&\sB_{0,0}+4i\chi\left(\ketbra{\Theta_-}{\Psi_+}+\ketbra{1,0}{1,0}-\ketbra{0,1}{0,1}\right)\,,\,\sB_{1,0}=\M\overline{\sB_{0,1}}\M
\end{align}
\end{widetext}
With this diagonal form of $\sQ\sL\sQ$, it is straightforward to compute $\left(\sQ\sL\sQ\right)^{-1}$. It consists in computing $\sB_{i,j}^{-1}$, $i,j=0,1$, which are $4\times4$ matrices. Moreover, since we are solely interested in quantities of the form $\sP\sL\sQ \left(\sQ\sL\sQ\right)^{-n}\sQ\sL\sP$ and $\sP\sL\sQ$ is of the form~\eqref{eq:2qubit_PLQ}, we only need to compute $\sB_{0,1}$.\\
To simplify this task, we define the unitary matrix
\begin{equation}
\sU=\frac{1}{\sqrt{2}}\ketbra{0,0}{\Psi_+}-\frac{1}{\sqrt{2}}\ketbra{1,1}{\Psi_-}+\ketbra{0,1}{0,1}+\ketbra{1,0}{1,0},
\end{equation}
and we compute the pseudo-inverse of $\tilde{\sB}_{0,1}=\sU\sB_{0,1}$. Multiplying by $\sU$ boils down to replacing the ket $\ket{\Psi_-}$ by $\ket{1,1}$ in~\eqref{App2:B01}, which implies that $\tilde{\sB}_{0,1}$ can be represented as a $3\times 4$ matrix in the standard basis "simplifying" the task of finding $\tilde{\sB}_{0,1}^{-1}$. To find the pseudo inverse of $\tilde{\sB}_{0,1}$, we simply solve the set of equations corresponding to:
\begin{equation}
\tilde{\sB}_{0,1}\tilde{\sB}_{0,1}^{-1}=\Pi_{\ran{\tilde{\sB}_{0,1}}}=\un^{(2B)}-\ketbra{0,0}{0,0}
\end{equation}
where $\Pi_{\ran{\tilde{\sB}_{0,1}}}$ is the hermitian projector to the range of $\tilde{\sB}_{0,1}$. If we define the following quantities
\begin{widetext}
\begin{align}
b_{11}&=- \frac{ \gamma \left(16 \chi^{2} + \gamma^{2} + 8\right)^{2}}{\sqrt{2} \left(2 i \chi \gamma \left(16 \chi^{2} + \gamma^{2} - 16\right) + \left(\gamma^{2} + 8\right) \left(16 \chi^{2} + \gamma^{2} + 8\right)\right) \left(16 \chi^{2} + \gamma^{2} + 4\right)}\\
b_{21}&=\frac{\sqrt{2} \left(16 \chi^{2} \gamma^{2} + 256 \chi^{2} - 4 i \chi \gamma \left(16 \chi^{2} + \gamma^{2}\right) + \gamma^{4} + 16 \gamma^{2} + 64\right)}{\left(2 i \chi \gamma \left(16 \chi^{2} + \gamma^{2} - 16\right) + \left(\gamma^{2} + 8\right) \left(16 \chi^{2} + \gamma^{2} + 8\right)\right) \left(16 \chi^{2} + \gamma^{2} + 4\right)}\\
b_{31}&= \frac{\sqrt{2} \left(16\chi^2\gamma^2- 4 i \chi \gamma \left(48 \chi^{2} + 3 \gamma^{2} + 16\right) + \left(32 \chi^{2} + \gamma^{2} + 8\right)^{2}\right)}{\left(2 i \chi \gamma \left(16 \chi^{2} + \gamma^{2} - 16\right) + \left(\gamma^{2} + 8\right) \left(16 \chi^{2} + \gamma^{2} + 8\right)\right) \left(16 \chi^{2} + \gamma^{2} + 4\right)}\\
b_{33}&=- \frac{512 \chi^{4} \gamma^{2} + 64 \chi^{2} \gamma^{4} + 576 \chi^{2} \gamma^{2} + 1024 \chi^{2} + 4 i \chi \gamma \left(16 \chi^{2} + \gamma^{2}\right)^{2} + 2 \gamma^{6} + 36 \gamma^{4} + 192 \gamma^{2} + 256}{\gamma \left(2 i \chi \gamma \left(16 \chi^{2} + \gamma^{2} - 16\right) + \left(\gamma^{2} + 8\right) \left(16 \chi^{2} + \gamma^{2} + 8\right)\right) \left(16 \chi^{2} + \gamma^{2} + 4\right)}\\
b_{22}&=b_{33}- \frac{32 \chi \left(8 \chi \left(16 \chi^{2} + \gamma^{2} + 4\right) - i \gamma \left(16 \chi^{2} + \gamma^{2} + 8\right)\right)}{\gamma \left(2 i \chi \gamma \left(16 \chi^{2} + \gamma^{2} - 16\right) + \left(\gamma^{2} + 8\right) \left(16 \chi^{2} + \gamma^{2} + 8\right)\right) \left(16 \chi^{2} + \gamma^{2} + 4\right)}\\
b_{12}&=b_{13}=\frac{2\sqrt{2}}{\gamma}b_{11}\,,\,b_{23}=\frac{2\sqrt{2}}{\gamma}b_{21}\,,\,b_{32}=\frac{2\sqrt{2}}{\gamma}b_{31}
\end{align}
\end{widetext}
and make the identification $\ket{1,1}\rightarrow\ket{1}$, $\ket{1,0}\rightarrow\ket{2}$ and $\ket{0,1}\rightarrow\ket{3}$, then:
\begin{equation}
\tilde{\sB}_{0,1}^{-1}=\sum_{i,j=1}^3b_{ij}\ketbra{i}{j}
\end{equation}
We can check that $\tilde{\sB}_{0,1}^{-1}$ verify all the Moore-Penrose conditions~\cite{Penrose1955}:
\begin{equation}
\begin{split}
\tilde{\sB}_{0,1}\tilde{\sB}_{0,1}^{-1}\tilde{\sB}_{0,1}=\tilde{\sB}_{0,1}&,
\tilde{\sB}_{0,1}^{-1}\tilde{\sB}_{0,1}\tilde{\sB}_{0,1}^{-1}=\tilde{\sB}_{0,1}^{-1}\\
\left(\tilde{\sB}_{0,1}\tilde{\sB}_{0,1}^{-1}\right)^\dagger=\tilde{\sB}_{0,1}\tilde{\sB}_{0,1}^{-1},&
\left(\tilde{\sB}_{0,1}^{-1}\tilde{\sB}_{0,1}\right)^\dagger=\tilde{\sB}_{0,1}^{-1}\tilde{\sB}_{0,1}
\end{split}
\end{equation}
Finally, we can easily see that the pseudo-inverse of ${\sB}_{0,1}$ is:
\begin{equation}
{\sB}_{0,1}^{-1}=\left(\sU^\dagger\tilde{\sB}_{0,1}\right)^{-1}=\tilde{\sB}_{0,1}^{-1}\sU
\end{equation}
with that, we have all the necessary ingredients to compute $\sL_0$ and $\sL_1$. Using Eqs.\eqref{eq:App2_PLPQ}, we find that:
\begin{equation}
\begin{split}
\sL_0&=\sP\sL\sP-\sP\sL\sQ\left(\sQ\sL\sQ\right)^{-1}\sQ\sL\sP \\
&=\ketbra{0,1}{0,1}\otimes\left[\sA_{0,1}-\sC_{0,1}\sB^{-1}_{0,1}\sD_{0,1}\right]\\
&+\ketbra{1,0}{1,0}\otimes\left[\sA_{1,0}-\sC_{1,0}\sB^{-1}_{1,0}\sD_{1,0}\right]
\end{split}
\end{equation}
\begin{equation}
\label{App_eq:L1_2qubit_final}
\begin{split}
\sL_1&=-\sP\sL\sQ\left(\sQ\sL\sQ\right)^{-2}\sQ\sL\sP \\
&=-\ketbra{0,1}{0,1}\otimes\left[\sC_{0,1}\sB^{-2}_{0,1}\sD_{0,1}\right]\\
&-\ketbra{1,0}{1,0}\otimes\left[\sC_{1,0}\sB^{-2}_{1,0}\sD_{1,0}\right]
\end{split}
\end{equation}
Since $\sP^{(B)}$ is a projector of rank $1$, we can write:
\begin{equation}
\begin{split}
\sA_{0,1}-\sC_{0,1}\sB^{-1}_{0,1}\sD_{0,1}&=\alpha_{0,1}\sP^{(B)}\\
\sA_{1,0}-\sC_{1,0}\sB^{-1}_{1,0}\sD_{1,0}&=\alpha_{1,0}\sP^{(B)}\\
\sC_{0,1}\sB^{-2}_{0,1}\sD_{0,1}&=\beta_{0,1}\sP^{(B)}\\
\sC_{1,0}\sB^{-2}_{1,0}\sD_{1,0}&=\beta_{1,0}\sP^{(B)}
\end{split}
\end{equation}
Because $\sP_B$ is of the form $\ket{\rho^{(B)}_{s_0}}\bra{\Psi_+}$, we find that:
\begin{equation}
\begin{split}
\alpha_{0,1}&\equiv \alpha=\frac{1}{2}\bra{\Psi_+}\left(\sA_{0,1}-\sC_{0,1}\sB^{-1}_{0,1}\sD_{0,1}\right)\ket{\Psi_+}\\
\beta_{0,1}&\equiv\beta=\frac{1}{2}\bra{\Psi_+}\left(\sC_{0,1}\sB^{-2}_{0,1}\sD_{0,1}\right)\ket{\Psi_+}
\end{split}
\end{equation}
where we have used the fact that $\braket{\Psi_+}{\rho^{(B)}_{s_0}}=1$ and $\braket{\Psi_+}{\Psi_+}=2$. We also have:
\begin{equation}
\begin{split}
\alpha_{1,0}&=\frac{1}{2}\bra{\Psi_+}\left(\sA_{1,0}-\sC_{1,0}\sB^{-1}_{1,0}\sD_{1,0}\right)\ket{\Psi_+}\\
&=\frac{1}{2}\bra{\Psi_+}\left(\M\overline{\sA_{0,1}}-\M\overline{\sC_{0,1}}\M\overline{\sB_{0,1}^{-1}}\overline{\sD_{0,1}}\right)\ket{\Psi_+}\\
&=\frac{1}{2}\bra{\Psi_+}\left(\overline{\sA_{0,1}}-\overline{\sC_{0,1}}\overline{\sB_{0,1}^{-1}}\overline{\sD_{0,1}}\right)\ket{\Psi_+}=\overline{\alpha}
\end{split}
\end{equation}
where we have used the fact that $\M^2=\un^{(2B)}$, $\overline{\sC_{0,1}}\M=\overline{\sC_{0,1}}$, and $\bra{\Psi_+}\M=\bra{\Psi_+}$. In a similar way we can also show that:
\begin{equation}
\beta_{1,0}=\overline{\beta}
\end{equation}
Let us define $\alpha=-\zeta+i\xi$ where $\zeta\geq 0$. A tedious calculation leads to:
\begin{widetext}
\begin{equation}
\label{App_alpha}
\begin{split}
\zeta&=\frac{128 \chi^{2} \gamma \left(\gamma^{2} + 8\right) \left(16 \chi^{2} + \gamma^{2} + 2\right)}{4 \chi^{2} \gamma^{2} \left(16 \chi^{2} + \gamma^{2} - 16\right)^{2} + \left(\gamma^{2} + 8\right)^{2} \left(16 \chi^{2} + \gamma^{2} + 8\right)^{2}}\\
\xi&=\frac{2 \chi \left(16 \chi^{2} + \gamma^{2}\right)}{16 \chi^{2} + \gamma^{2} + 8}+\frac{256 \chi^{3} \gamma^{2} \left(16 \chi^{2} + \gamma^{2} - 16\right) \left(16 \chi^{2} + \gamma^{2} + 2\right)}{\left(4 \chi^{2} \gamma^{2} \left(16 \chi^{2} + \gamma^{2} - 16\right)^{2} + \left(\gamma^{2} + 8\right)^{2} \left(16 \chi^{2} + \gamma^{2} + 8\right)^{2}\right) \left(16 \chi^{2} + \gamma^{2} + 8\right)}
\end{split}
\end{equation}
and
\[\beta=\frac{x_1+iy_1}{x_2+iy_2}\]
where
\begin{equation}
\begin{split}
x_1=&\chi^{2} \left(49152 \chi^{4} \gamma^{2} - 262144 \chi^{4} + 6144 \chi^{2} \gamma^{4} + 2048 \chi^{2} \gamma^{2} - 131072 \chi^{2} + 192 \gamma^{6} + 1152 \gamma^{4} - 5120 \gamma^{2} - 16384\right)\\
y_1=&256 \chi^{3} \gamma \left(16 \chi^{2} + \gamma^{2} + 4\right) \left(16 \chi^{2} + \gamma^{2} + 8\right)\\
x_2=&\left(16 \chi^{2} + \gamma^{2} + 4\right) \left(- 32 \chi^{3} \gamma + 16 \chi^{2} \gamma^{2} + 128 \chi^{2} - 2 \chi \gamma^{3} + 32 \chi \gamma + \gamma^{4} + 16 \gamma^{2} + 64\right) \\
&\times\left(32 \chi^{3} \gamma + 16 \chi^{2} \gamma^{2} + 128 \chi^{2} + 2 \chi \gamma^{3} - 32 \chi \gamma + \gamma^{4} + 16 \gamma^{2} + 64\right)\\
y_2=&4 \chi \gamma \left(\gamma^{2} + 8\right) \left(16 \chi^{2} + \gamma^{2} - 16\right) \left(16 \chi^{2} + \gamma^{2} + 4\right) \left(16 \chi^{2} + \gamma^{2} + 8\right)
\end{split}
\end{equation}
\end{widetext}
Finally, it is straightforward to calculate $\left(\un-\sL_1\right)^{-1}$ exactly , given that $\sL_1$~\eqref{App_eq:L1_2qubit_final} is block diagonal:
\begin{equation}
\begin{split}
\left(\un-\sL_1\right)^{-1}&=\ketbra{0,1}{0,1}\otimes\left[\frac{1}{1+\beta}\sP_B+\sQ_B\right]\\
&+\ketbra{1,0}{1,0}\otimes\left[\frac{1}{1+\overline{\beta}}\sP_B+\sQ_B\right]
\\
&+\left[\ketbra{0,0}{0,0}+\ketbra{1,1}{1,1}\right]\otimes \un^{(2B)}
\end{split}
\end{equation}
Taking the fact $\sQ_B\rho^{(B)}_{s_0}=0$ into account and defining the modified initial state of the qubit $A$ as:
\begin{equation}
\ket{\tilde{\rho}^{(A)}_0}=\rho_{0,0}\ket{0,0}+\rho_{1,1}\ket{1,1}+\frac{\rho_{0,1}}{1+\beta}\ket{0,1}+\frac{\rho_{1,0}}{1+\overline{\beta}}\ket{1,0},
\end{equation}
where $\rho_{0,0}, \rho_{1,1}$ represent population in the state $\rho^{(A)}_0$ while $\rho_{0,1}, \rho_{1,0}$ represent initial coherences, we can simplify~\eqref{eq:approxDynamics} to describe the dynamics of the slow qubit as:
\begin{equation}
\ket{\rho^A(t)}=e^{\tilde{L}_0t}\ket{\tilde{\rho}^A_0},
\end{equation}
where $\tilde{L}_0=\frac{\alpha}{1+\beta}\ketbra{0,1}{0,1}+\frac{\overline{\alpha}}{1+\overline{\beta}}\ketbra{1,0}{1,0}$. The evolution operator $U(t)=e^{\tilde{L}_0t}$ is simple to calculate:
\begin{equation}
\begin{split}
U(t)&=e^{-\zeta' t+i\xi' t}\ketbra{0,1}{0,1}+e^{-\zeta' t-i\xi' t}\ketbra{1,0}{1,0}\\
&+\ketbra{0,0}{0,0}+\ketbra{1,1}{1,1}
\end{split}
\end{equation}
where we have defined
\begin{equation}
\zeta'=-\Re{\frac{\alpha}{1+\beta}}\,,\,\xi'=\Im{\frac{\alpha}{1+\beta}}
\end{equation}
\section{open Rabi model}
\label{App.SB}
In this section, we carry out all the calculations needed to adiabatically eliminate a fast qubit interacting with a slow cavity mode according to the open Rabi model:
\begin{equation}
\begin{split}
\sL\left(\rho\right)=&-i\frac{1}{\eta}\commut{\sigma_z}{\rho}+\Gamma\sD_{\sigma_-}\left(\rho\right)-i{\eta}\commut{ a^\dagger a}\left(\rho\right)\\& +\kappa \sD_a\left(\rho\right)-ig\commut{\left(a+a^\dagger\right)\otimes\sigma_x}{\rho}
\end{split}
\end{equation}
The first step is to write $\sL$ in the super-operator representation:
\begin{widetext}
\begin{equation}
\begin{split}
\sL=&-i\eta \left(\un^{(A)}\otimes a^{\dagger}a \otimes \un^{(2B)}-a^{\dagger}a\otimes\un^{(A)}\otimes  \un^{(2B)}\right)
-\dfrac{i}{\eta}\left(\un^{(2A)}\otimes \un^{(B)}\otimes \sigma_z-\un^{(2A)}\otimes  \sigma_z\otimes\un^{(B)}\right)\\
& -ig\left(\un^{(A)}\otimes\left(a+a^{\dagger}\right)\otimes \un^{(B)}\otimes \sigma_x-\left(a+a^{\dagger}\right)\otimes\un^{(A)}\otimes  \sigma_x\otimes\un^{(B)}\right)\nonumber\\
&+\Gamma\un^{(2A)}\otimes \sigma_-\otimes\sigma_--\dfrac{\Gamma}{2}\left[\un^{(2A)}\otimes \sigma_+\sigma_-\otimes\un^{(B)}+\un^{(2A)}\otimes \un^{(B)}\otimes \sigma_+\sigma_-\right]\\
&+\kappa a\otimes a \otimes \un^{(2B)}-\dfrac{\kappa}{2}\left[a^{\dagger}a\otimes \un^{(A)}\otimes\un^{(2B)}+\un^{(A)}\otimes a^{\dagger}a\otimes \un^{(2B)}\right]
\end{split}
\end{equation}
If we define:
\begin{equation}
\sP_B=\kket{\rho^B_f}\bbra{\un^{(B)}}\,,\,\sQ_B=\un^{(2B)}-\sP_B=\ketbra{1,1}{1,1}-\ketbra{0,0}{1,1}+\ketbra{0,1}{0,1}+\ketbra{1,0}{1,0}
\end{equation}
and
\begin{equation}
\sP=\un^{(2A)}\otimes \sP_B \,,\,\sQ= \un^{(2A)}\otimes \sQ_B
\end{equation}
then we can compute the needed quantities for $\sL_0$
\begin{eqnarray}
\sP\sL\sP=&\left[-i\eta \left(\un^{(A)}\otimes a^{\dagger}a -a^{\dagger}a\otimes\un^{(A)}\right)+\kappa a\otimes a -\dfrac{\kappa}{2}\left[a^{\dagger}a\otimes \un^{(A)}+\un^{(A)}\otimes a^{\dagger}a \right]\right]\otimes P_B\nonumber\\
\sP\sL\sQ=&-ig\left[\un^{(A)}\otimes\left(a+a^{\dagger}\right)-\left(a+a^{\dagger}\right)\otimes\un^{(A)}\right]\otimes\ketbra{0,0}{\Phi_+}\nonumber\\
\sQ\sL\sP=&-ig\left(\un^{(A)}\otimes\left(a+a^{\dagger}\right)\otimes \ketbra{0,1}{\Psi_+}-\left(a+a^{\dagger}\right)\otimes\un^{(A)}\otimes  \ketbra{1,0}{\Psi_+}\right)\nonumber\\
\sQ\sL_B\sQ&=\un^{(2A)}\otimes\left[-\Gamma \ketbra{1,1}{1,1}+\Gamma \ketbra{0,0}{1,1}-\left(\dfrac{\Gamma}{2}+\dfrac{2i}{\eta}\right)\ketbra{0,1}{0,1}-\left(\dfrac{\Gamma}{2}-\dfrac{2i}{\eta}\right)\ketbra{1,0}{1,0}\right]
\end{eqnarray}

$\sQ\sL_B\sQ$ represents the dominant term of $\sQ\sL\sQ$ which can be inverted quite easily:
\begin{equation}
Q\sL_BQ^{-1}=\un^{(2A)}\otimes\left[-\dfrac{1}{2\Gamma} \ketbra{1,1}{1,1}+\dfrac{1}{2\Gamma} \ketbra{1,1}{0,0}-\dfrac{1}{\left(\dfrac{\Gamma}{2}+\dfrac{2i}{\eta}\right)}\ketbra{0,1}{0,1}-\dfrac{1}{\left(\dfrac{\Gamma}{2}-\dfrac{2i}{\eta}\right)}\ketbra{1,0}{1,0}\right]
\end{equation}

Hence we can easily calculate $\sL_0$ to be
\begin{eqnarray}
\sL_0=&\left[-i\eta \left(\un^{(A)}\otimes a^{\dagger}a -a^{\dagger}a\otimes\un^{(A)}\right)+\kappa a\otimes a -\dfrac{\kappa}{2}\left[a^{\dagger}a\otimes \un^{(A)}+\un^{(A)}\otimes a^{\dagger}a\right]\right.\nonumber\\
&\left.-g^2\dfrac{\dfrac{\Gamma}{2}-\dfrac{2i}{\eta}}{\dfrac{\Gamma^2}{4}+\dfrac{4}{\eta^2}}\left[\un^{(A)}\otimes\left(a+a^{\dagger}\right)^2-\left(a+a^{\dagger}\right)\otimes\left(a+a^{\dagger}\right)\right]\right.\nonumber\\
&\left.-g^2\dfrac{\dfrac{\Gamma}{2}+\dfrac{2i}{\eta}}{\dfrac{\Gamma^2}{4}+\dfrac{4}{\eta^2}}\left[\left(a+a^{\dagger}\right)^2\otimes \un^{(A)}-\left(a+a^{\dagger}\right)\otimes\left(a+a^{\dagger}\right)\right]\right]\otimes \sP_B
\end{eqnarray}
\end{widetext}
From which we can deduce the reduced dynamics governing the evolution of the slow system to be:
\begin{equation}
\sL_0\left(\rho^{(A)}\right)=-i\commut{H^{(A)}}{\rho^{(A)}}+\kappa\sD_{a}\left(\rho^{(A)}\right)+\dfrac{4g^2\eta^2\Gamma}{\Gamma^2\eta^2+16}\sD_{\left(a+a^{\dagger}\right)}\left(\rho^{(A)}\right)
\end{equation}
where we have defined:
\begin{equation}
H^{(A)}=\eta a^{\dagger}{a}-\dfrac{4g^2\eta}{\Gamma^2\eta^2+16}\left(a+a^{\dagger}\right)^2
\end{equation}
\bibliography{adiabElimin}

\begin{thebibliography}{38}%
\makeatletter
\providecommand \@ifxundefined [1]{%
 \@ifx{#1\undefined}
}%
\providecommand \@ifnum [1]{%
 \ifnum #1\expandafter \@firstoftwo
 \else \expandafter \@secondoftwo
 \fi
}%
\providecommand \@ifx [1]{%
 \ifx #1\expandafter \@firstoftwo
 \else \expandafter \@secondoftwo
 \fi
}%
\providecommand \natexlab [1]{#1}%
\providecommand \enquote  [1]{``#1''}%
\providecommand \bibnamefont  [1]{#1}%
\providecommand \bibfnamefont [1]{#1}%
\providecommand \citenamefont [1]{#1}%
\providecommand \href@noop [0]{\@secondoftwo}%
\providecommand \href [0]{\begingroup \@sanitize@url \@href}%
\providecommand \@href[1]{\@@startlink{#1}\@@href}%
\providecommand \@@href[1]{\endgroup#1\@@endlink}%
\providecommand \@sanitize@url [0]{\catcode `\\12\catcode `\$12\catcode
  `\&12\catcode `\#12\catcode `\^12\catcode `\_12\catcode `\%12\relax}%
\providecommand \@@startlink[1]{}%
\providecommand \@@endlink[0]{}%
\providecommand \url  [0]{\begingroup\@sanitize@url \@url }%
\providecommand \@url [1]{\endgroup\@href {#1}{\urlprefix }}%
\providecommand \urlprefix  [0]{URL }%
\providecommand \Eprint [0]{\href }%
\providecommand \doibase [0]{http://dx.doi.org/}%
\providecommand \selectlanguage [0]{\@gobble}%
\providecommand \bibinfo  [0]{\@secondoftwo}%
\providecommand \bibfield  [0]{\@secondoftwo}%
\providecommand \translation [1]{[#1]}%
\providecommand \BibitemOpen [0]{}%
\providecommand \bibitemStop [0]{}%
\providecommand \bibitemNoStop [0]{.\EOS\space}%
\providecommand \EOS [0]{\spacefactor3000\relax}%
\providecommand \BibitemShut  [1]{\csname bibitem#1\endcsname}%
\let\auto@bib@innerbib\@empty
\bibitem [{\citenamefont {Haken}(1975)}]{Haken1975}%
  \BibitemOpen
  \bibfield  {author} {\bibinfo {author} {\bibfnamefont {H.}~\bibnamefont
  {Haken}},\ }\href@noop {} {\bibfield  {journal} {\bibinfo  {journal} {Z
  Physik B}\ }\textbf {\bibinfo {volume} {20}},\ \bibinfo {pages} {413}
  (\bibinfo {year} {1975})}\BibitemShut {NoStop}%
\bibitem [{\citenamefont {Haken}(1977)}]{Haken1977}%
  \BibitemOpen
  \bibfield  {author} {\bibinfo {author} {\bibnamefont {Haken}},\ }\href@noop
  {} {\emph {\bibinfo {title} {Synergetics--An introduction}}}\ (\bibinfo
  {publisher} {Springer Berlin},\ \bibinfo {year} {1977})\BibitemShut {NoStop}%
\bibitem [{\citenamefont {Lax}(1967)}]{Lax1967}%
  \BibitemOpen
  \bibfield  {author} {\bibinfo {author} {\bibfnamefont {M.}~\bibnamefont
  {Lax}},\ }\href@noop {} {\bibfield  {journal} {\bibinfo  {journal} {Phys.
  Rev.}\ }\textbf {\bibinfo {volume} {157}},\ \bibinfo {pages} {213} (\bibinfo
  {year} {1967})}\BibitemShut {NoStop}%
\bibitem [{\citenamefont {Cohen-Tannoudji}(1992)}]{Cohen-tannoudji1992}%
  \BibitemOpen
  \bibfield  {author} {\bibinfo {author} {\bibfnamefont {C.}~\bibnamefont
  {Cohen-Tannoudji}},\ }\href@noop {} {\bibfield  {journal} {\bibinfo
  {journal} {Physics Reports}\ }\textbf {\bibinfo {volume} {219}},\ \bibinfo
  {pages} {153} (\bibinfo {year} {1992})}\BibitemShut {NoStop}%
\bibitem [{\citenamefont {Paulisch}\ \emph {et~al.}(2014)\citenamefont
  {Paulisch}, \citenamefont {Rui}, \citenamefont {Ng},\ and\ \citenamefont
  {Englert}}]{Paulisch_2014}%
  \BibitemOpen
  \bibfield  {author} {\bibinfo {author} {\bibfnamefont {V.}~\bibnamefont
  {Paulisch}}, \bibinfo {author} {\bibfnamefont {H.}~\bibnamefont {Rui}},
  \bibinfo {author} {\bibfnamefont {H.~K.}\ \bibnamefont {Ng}}, \ and\ \bibinfo
  {author} {\bibfnamefont {B.-G.}\ \bibnamefont {Englert}},\ }\href@noop {}
  {\bibfield  {journal} {\bibinfo  {journal} {Eur. Phys. J. Plus}\ }\textbf
  {\bibinfo {volume} {129}},\ \bibinfo {pages} {12} (\bibinfo {year}
  {2014})}\BibitemShut {NoStop}%
\bibitem [{\citenamefont {Brion}\ \emph {et~al.}(2007)\citenamefont {Brion},
  \citenamefont {Pedersen},\ and\ \citenamefont {M{\o}lmer}}]{Brion2007}%
  \BibitemOpen
  \bibfield  {author} {\bibinfo {author} {\bibfnamefont {E.}~\bibnamefont
  {Brion}}, \bibinfo {author} {\bibfnamefont {L.~H.}\ \bibnamefont {Pedersen}},
  \ and\ \bibinfo {author} {\bibfnamefont {K.}~\bibnamefont {M{\o}lmer}},\
  }\href {"https://doi.org/10.1088%2F1751-8113%2F40%2F5%2F011"} {\bibfield
  {journal} {\bibinfo  {journal} {J. Phys. A: Math. Theor.}\ }\textbf {\bibinfo
  {volume} {40}},\ \bibinfo {pages} {1033} (\bibinfo {year}
  {2007})}\BibitemShut {NoStop}%
\bibitem [{\citenamefont {You}\ \emph {et~al.}(2003)\citenamefont {You},
  \citenamefont {Yi},\ and\ \citenamefont {Su}}]{You2003}%
  \BibitemOpen
  \bibfield  {author} {\bibinfo {author} {\bibfnamefont {L.}~\bibnamefont
  {You}}, \bibinfo {author} {\bibfnamefont {X.~X.}\ \bibnamefont {Yi}}, \ and\
  \bibinfo {author} {\bibfnamefont {X.~H.}\ \bibnamefont {Su}},\ }\href@noop {}
  {\bibfield  {journal} {\bibinfo  {journal} {Phys. Rev. A}\ }\textbf {\bibinfo
  {volume} {67}},\ \bibinfo {pages} {032308} (\bibinfo {year}
  {2003})}\BibitemShut {NoStop}%
\bibitem [{\citenamefont {Nagy}\ \emph {et~al.}(2010)\citenamefont {Nagy},
  \citenamefont {K{\'o}nya}, \citenamefont {Szirmai},\ and\ \citenamefont
  {Domokos}}]{Nagy2010}%
  \BibitemOpen
  \bibfield  {author} {\bibinfo {author} {\bibfnamefont {D.}~\bibnamefont
  {Nagy}}, \bibinfo {author} {\bibfnamefont {G.}~\bibnamefont {K{\'o}nya}},
  \bibinfo {author} {\bibfnamefont {G.}~\bibnamefont {Szirmai}}, \ and\
  \bibinfo {author} {\bibfnamefont {P.}~\bibnamefont {Domokos}},\ }\href
  {https://link.aps.org/doi/10.1103/PhysRevLett.104.130401} {\bibfield
  {journal} {\bibinfo  {journal} {Phys. Rev. Lett.}\ }\textbf {\bibinfo
  {volume} {104}},\ \bibinfo {pages} {130401} (\bibinfo {year}
  {2010})}\BibitemShut {NoStop}%
\bibitem [{\citenamefont {Douglas}\ \emph {et~al.}(2015)\citenamefont
  {Douglas}, \citenamefont {Habibian}, \citenamefont {Hung}, \citenamefont
  {Gorshkov}, \citenamefont {Kimble},\ and\ \citenamefont
  {Chang}}]{Douglas2015}%
  \BibitemOpen
  \bibfield  {author} {\bibinfo {author} {\bibfnamefont {J.~S.}\ \bibnamefont
  {Douglas}}, \bibinfo {author} {\bibfnamefont {H.}~\bibnamefont {Habibian}},
  \bibinfo {author} {\bibfnamefont {C.-L.}\ \bibnamefont {Hung}}, \bibinfo
  {author} {\bibfnamefont {A.~V.}\ \bibnamefont {Gorshkov}}, \bibinfo {author}
  {\bibfnamefont {H.~J.}\ \bibnamefont {Kimble}}, \ and\ \bibinfo {author}
  {\bibfnamefont {D.~E.}\ \bibnamefont {Chang}},\ }\href@noop {} {\bibfield
  {journal} {\bibinfo  {journal} {Nature Photonics}\ }\textbf {\bibinfo
  {volume} {9}},\ \bibinfo {pages} {326} (\bibinfo {year} {2015})}\BibitemShut
  {NoStop}%
\bibitem [{\citenamefont {Sinatra}\ \emph {et~al.}(1995)\citenamefont
  {Sinatra}, \citenamefont {Castelli}, \citenamefont {Lugiato}, \citenamefont
  {Grangier},\ and\ \citenamefont {Poizat}}]{Sinatra1995}%
  \BibitemOpen
  \bibfield  {author} {\bibinfo {author} {\bibfnamefont {A.}~\bibnamefont
  {Sinatra}}, \bibinfo {author} {\bibfnamefont {F.}~\bibnamefont {Castelli}},
  \bibinfo {author} {\bibfnamefont {L.~A.}\ \bibnamefont {Lugiato}}, \bibinfo
  {author} {\bibfnamefont {P.}~\bibnamefont {Grangier}}, \ and\ \bibinfo
  {author} {\bibfnamefont {J.~P.}\ \bibnamefont {Poizat}},\ }\href@noop {}
  {\bibfield  {journal} {\bibinfo  {journal} {Quantum Semiclass. Opt.}\
  }\textbf {\bibinfo {volume} {7}},\ \bibinfo {pages} {405} (\bibinfo {year}
  {1995})}\BibitemShut {NoStop}%
\bibitem [{\citenamefont {Azouit}\ \emph {et~al.}(2016)\citenamefont {Azouit},
  \citenamefont {Sarlette},\ and\ \citenamefont {Rouchon}}]{Azouit2016}%
  \BibitemOpen
  \bibfield  {author} {\bibinfo {author} {\bibfnamefont {R.}~\bibnamefont
  {Azouit}}, \bibinfo {author} {\bibfnamefont {A.}~\bibnamefont {Sarlette}}, \
  and\ \bibinfo {author} {\bibfnamefont {P.}~\bibnamefont {Rouchon}},\
  }\href@noop {} {\bibfield  {journal} {\bibinfo  {journal} {arXiv:1603.04630
  [quant-ph]}\ } (\bibinfo {year} {2016})},\ \bibinfo {note} {arXiv:
  1603.04630}\BibitemShut {NoStop}%
\bibitem [{\citenamefont {Azouit}\ \emph
  {et~al.}(2017{\natexlab{a}})\citenamefont {Azouit}, \citenamefont {Chittaro},
  \citenamefont {Sarlette},\ and\ \citenamefont {Rouchon}}]{Azouit2017}%
  \BibitemOpen
  \bibfield  {author} {\bibinfo {author} {\bibfnamefont {R.}~\bibnamefont
  {Azouit}}, \bibinfo {author} {\bibfnamefont {F.}~\bibnamefont {Chittaro}},
  \bibinfo {author} {\bibfnamefont {A.}~\bibnamefont {Sarlette}}, \ and\
  \bibinfo {author} {\bibfnamefont {P.}~\bibnamefont {Rouchon}},\ }\href@noop
  {} {\bibfield  {journal} {\bibinfo  {journal} {Quantum Sci. Technol.}\
  }\textbf {\bibinfo {volume} {2}},\ \bibinfo {pages} {044011} (\bibinfo {year}
  {2017}{\natexlab{a}})}\BibitemShut {NoStop}%
\bibitem [{\citenamefont {Azouit}(2017)}]{AzouitThesis2017}%
  \BibitemOpen
  \bibfield  {author} {\bibinfo {author} {\bibfnamefont {R.}~\bibnamefont
  {Azouit}},\ }\emph {\bibinfo {title} {Adiabatic elimination for open quantum
  systems}},\ \href@noop {} {Ph.D. thesis},\ \bibinfo  {school} {PSL Research
  University} (\bibinfo {year} {2017})\BibitemShut {NoStop}%
\bibitem [{\citenamefont {Azouit}\ \emph
  {et~al.}(2017{\natexlab{b}})\citenamefont {Azouit}, \citenamefont {Chittaro},
  \citenamefont {Sarlette},\ and\ \citenamefont
  {Rouchon}}]{Azouit_structure-preserving_2017}%
  \BibitemOpen
  \bibfield  {author} {\bibinfo {author} {\bibfnamefont {R.}~\bibnamefont
  {Azouit}}, \bibinfo {author} {\bibfnamefont {F.}~\bibnamefont {Chittaro}},
  \bibinfo {author} {\bibfnamefont {A.}~\bibnamefont {Sarlette}}, \ and\
  \bibinfo {author} {\bibfnamefont {P.}~\bibnamefont {Rouchon}},\ }\href@noop
  {} {\bibfield  {journal} {\bibinfo  {journal} {IFAC-PapersOnLine}\ }\bibinfo
  {series} {20th {IFAC} {World} {Congress}},\ \textbf {\bibinfo {volume}
  {50}},\ \bibinfo {pages} {13026} (\bibinfo {year}
  {2017}{\natexlab{b}})}\BibitemShut {NoStop}%
\bibitem [{\citenamefont {Sarlette}\ \emph {et~al.}(2020)\citenamefont
  {Sarlette}, \citenamefont {Rouchon}, \citenamefont {Essig}, \citenamefont
  {Ficheux},\ and\ \citenamefont {Huard}}]{Sarlette2020}%
  \BibitemOpen
  \bibfield  {author} {\bibinfo {author} {\bibfnamefont {A.}~\bibnamefont
  {Sarlette}}, \bibinfo {author} {\bibfnamefont {P.}~\bibnamefont {Rouchon}},
  \bibinfo {author} {\bibfnamefont {A.}~\bibnamefont {Essig}}, \bibinfo
  {author} {\bibfnamefont {Q.}~\bibnamefont {Ficheux}}, \ and\ \bibinfo
  {author} {\bibfnamefont {B.}~\bibnamefont {Huard}},\ }\href
  {http://arxiv.org/abs/2001.02550} {\  (\bibinfo {year} {2020})},\ \Eprint
  {http://arxiv.org/abs/2001.02550} {arXiv:2001.02550} \BibitemShut {NoStop}%
\bibitem [{\citenamefont {Mirrahimi}\ and\ \citenamefont
  {Rouchon}(2009)}]{Mirrahimi2009}%
  \BibitemOpen
  \bibfield  {author} {\bibinfo {author} {\bibfnamefont {M.}~\bibnamefont
  {Mirrahimi}}\ and\ \bibinfo {author} {\bibfnamefont {P.}~\bibnamefont
  {Rouchon}},\ }\href@noop {} {\bibfield  {journal} {\bibinfo  {journal} {IEEE
  Transactions on Automatic Control}\ }\textbf {\bibinfo {volume} {54}},\
  \bibinfo {pages} {1325} (\bibinfo {year} {2009})}\BibitemShut {NoStop}%
\bibitem [{\citenamefont {Reiter}\ and\ \citenamefont
  {S\o{}rensen}(2012)}]{Reiter2012}%
  \BibitemOpen
  \bibfield  {author} {\bibinfo {author} {\bibfnamefont {F.}~\bibnamefont
  {Reiter}}\ and\ \bibinfo {author} {\bibfnamefont {A.~S.}\ \bibnamefont
  {S\o{}rensen}},\ }\href@noop {} {\bibfield  {journal} {\bibinfo  {journal}
  {Phys. Rev. A}\ }\textbf {\bibinfo {volume} {85}},\ \bibinfo {pages} {032111}
  (\bibinfo {year} {2012})}\BibitemShut {NoStop}%
\bibitem [{\citenamefont {Kessler}(2012)}]{Kessler2012}%
  \BibitemOpen
  \bibfield  {author} {\bibinfo {author} {\bibfnamefont {E.~M.}\ \bibnamefont
  {Kessler}},\ }\href {\doibase 10.1103/PhysRevA.86.012126} {\bibfield
  {journal} {\bibinfo  {journal} {Phys. Rev. A}\ }\textbf {\bibinfo {volume}
  {86}},\ \bibinfo {pages} {012126} (\bibinfo {year} {2012})}\BibitemShut
  {NoStop}%
\bibitem [{\citenamefont {Lin}\ \emph {et~al.}(2013)\citenamefont {Lin},
  \citenamefont {Gaebler}, \citenamefont {Reiter}, \citenamefont {Tan},
  \citenamefont {Bowler}, \citenamefont {S{\o}rensen}, \citenamefont
  {Leibfried},\ and\ \citenamefont {Wineland}}]{Lin2013}%
  \BibitemOpen
  \bibfield  {author} {\bibinfo {author} {\bibfnamefont {Y.}~\bibnamefont
  {Lin}}, \bibinfo {author} {\bibfnamefont {J.~P.}\ \bibnamefont {Gaebler}},
  \bibinfo {author} {\bibfnamefont {F.}~\bibnamefont {Reiter}}, \bibinfo
  {author} {\bibfnamefont {T.~R.}\ \bibnamefont {Tan}}, \bibinfo {author}
  {\bibfnamefont {R.}~\bibnamefont {Bowler}}, \bibinfo {author} {\bibfnamefont
  {A.~S.}\ \bibnamefont {S{\o}rensen}}, \bibinfo {author} {\bibfnamefont
  {D.}~\bibnamefont {Leibfried}}, \ and\ \bibinfo {author} {\bibfnamefont
  {D.~J.}\ \bibnamefont {Wineland}},\ }\href
  {https://doi.org/10.1038/nature12801} {\bibfield  {journal} {\bibinfo
  {journal} {Nature}\ }\textbf {\bibinfo {volume} {504}},\ \bibinfo {pages}
  {415} (\bibinfo {year} {2013})}\BibitemShut {NoStop}%
\bibitem [{\citenamefont {Albert}\ \emph {et~al.}(2019)\citenamefont {Albert},
  \citenamefont {Noh},\ and\ \citenamefont {Reiterr}}]{Albert2019}%
  \BibitemOpen
  \bibfield  {author} {\bibinfo {author} {\bibfnamefont {V.}~\bibnamefont
  {Albert}}, \bibinfo {author} {\bibfnamefont {K.}~\bibnamefont {Noh}}, \ and\
  \bibinfo {author} {\bibfnamefont {F.}~\bibnamefont {Reiterr}},\ }\href@noop
  {} {\bibfield  {journal} {\bibinfo  {journal} {arXiv:1809.07324}\ } (\bibinfo
  {year} {2019})}\BibitemShut {NoStop}%
\bibitem [{\citenamefont {Pastawski}\ \emph {et~al.}(2011)\citenamefont
  {Pastawski}, \citenamefont {Clemente},\ and\ \citenamefont
  {Cirac}}]{Pastawski2011}%
  \BibitemOpen
  \bibfield  {author} {\bibinfo {author} {\bibfnamefont {F.}~\bibnamefont
  {Pastawski}}, \bibinfo {author} {\bibfnamefont {L.}~\bibnamefont {Clemente}},
  \ and\ \bibinfo {author} {\bibfnamefont {J.~I.}\ \bibnamefont {Cirac}},\
  }\href {\doibase 10.1103/PhysRevA.83.012304} {\bibfield  {journal} {\bibinfo
  {journal} {Phys. Rev. A}\ }\textbf {\bibinfo {volume} {83}},\ \bibinfo
  {pages} {012304} (\bibinfo {year} {2011})}\BibitemShut {NoStop}%
\bibitem [{\citenamefont {\ifmmode~\check{C}\else \v{C}\fi{}ernot\'{\i}k}\
  \emph {et~al.}(2015)\citenamefont {\ifmmode~\check{C}\else
  \v{C}\fi{}ernot\'{\i}k}, \citenamefont {Vasilyev},\ and\ \citenamefont
  {Hammerer}}]{Cernotik2015}%
  \BibitemOpen
  \bibfield  {author} {\bibinfo {author} {\bibfnamefont {O.~c.~v.}\
  \bibnamefont {\ifmmode~\check{C}\else \v{C}\fi{}ernot\'{\i}k}}, \bibinfo
  {author} {\bibfnamefont {D.~V.}\ \bibnamefont {Vasilyev}}, \ and\ \bibinfo
  {author} {\bibfnamefont {K.}~\bibnamefont {Hammerer}},\ }\href {\doibase
  10.1103/PhysRevA.92.012124} {\bibfield  {journal} {\bibinfo  {journal} {Phys.
  Rev. A}\ }\textbf {\bibinfo {volume} {92}},\ \bibinfo {pages} {012124}
  (\bibinfo {year} {2015})}\BibitemShut {NoStop}%
\bibitem [{\citenamefont {Minganti}\ \emph {et~al.}(2018)\citenamefont
  {Minganti}, \citenamefont {Biella}, \citenamefont {Bartolo},\ and\
  \citenamefont {Ciuti}}]{Ciuti2018}%
  \BibitemOpen
  \bibfield  {author} {\bibinfo {author} {\bibfnamefont {F.}~\bibnamefont
  {Minganti}}, \bibinfo {author} {\bibfnamefont {A.}~\bibnamefont {Biella}},
  \bibinfo {author} {\bibfnamefont {N.}~\bibnamefont {Bartolo}}, \ and\
  \bibinfo {author} {\bibfnamefont {C.}~\bibnamefont {Ciuti}},\ }\href@noop {}
  {\bibfield  {journal} {\bibinfo  {journal} {Physical Review A}\ }\textbf
  {\bibinfo {volume} {98}},\ \bibinfo {pages} {042118} (\bibinfo {year}
  {2018})}\BibitemShut {NoStop}%
\bibitem [{\citenamefont {Finkelstein-Shapiro}\ \emph
  {et~al.}(2020)\citenamefont {Finkelstein-Shapiro}, \citenamefont {Viennot},
  \citenamefont {Saideh}, \citenamefont {Hansen}, \citenamefont {Pullerits},\
  and\ \citenamefont {Keller}}]{Finkelstein-Shapiro2020}%
  \BibitemOpen
  \bibfield  {author} {\bibinfo {author} {\bibfnamefont {D.}~\bibnamefont
  {Finkelstein-Shapiro}}, \bibinfo {author} {\bibfnamefont {D.}~\bibnamefont
  {Viennot}}, \bibinfo {author} {\bibfnamefont {I.}~\bibnamefont {Saideh}},
  \bibinfo {author} {\bibfnamefont {T.}~\bibnamefont {Hansen}}, \bibinfo
  {author} {\bibfnamefont {T.~o.}\ \bibnamefont {Pullerits}}, \ and\ \bibinfo
  {author} {\bibfnamefont {A.}~\bibnamefont {Keller}},\ }\href@noop {}
  {\bibfield  {journal} {\bibinfo  {journal} {Phys. Rev. A}\ }\textbf {\bibinfo
  {volume} {101}},\ \bibinfo {pages} {042102} (\bibinfo {year}
  {2020})}\BibitemShut {NoStop}%
\bibitem [{\citenamefont {Lesanovsky}\ and\ \citenamefont
  {Garrahan}(2013)}]{Lesanovsky2013}%
  \BibitemOpen
  \bibfield  {author} {\bibinfo {author} {\bibfnamefont {I.}~\bibnamefont
  {Lesanovsky}}\ and\ \bibinfo {author} {\bibfnamefont {J.~P.}\ \bibnamefont
  {Garrahan}},\ }\href@noop {} {\bibfield  {journal} {\bibinfo  {journal}
  {Phys. Rev. Lett.}\ }\textbf {\bibinfo {volume} {111}},\ \bibinfo {pages}
  {215305} (\bibinfo {year} {2013})}\BibitemShut {NoStop}%
\bibitem [{\citenamefont {Marcuzzi}\ \emph {et~al.}(2014)\citenamefont
  {Marcuzzi}, \citenamefont {Schick}, \citenamefont {Olmos},\ and\
  \citenamefont {Lesanovsky}}]{Marcuzzi2014}%
  \BibitemOpen
  \bibfield  {author} {\bibinfo {author} {\bibfnamefont {M.}~\bibnamefont
  {Marcuzzi}}, \bibinfo {author} {\bibfnamefont {J.}~\bibnamefont {Schick}},
  \bibinfo {author} {\bibfnamefont {B.}~\bibnamefont {Olmos}}, \ and\ \bibinfo
  {author} {\bibfnamefont {I.}~\bibnamefont {Lesanovsky}},\ }\href {\doibase
  10.1088/1751-8113/47/48/482001} {\bibfield  {journal} {\bibinfo  {journal}
  {Journal of Physics A: Mathematical and Theoretical}\ }\textbf {\bibinfo
  {volume} {47}},\ \bibinfo {pages} {482001} (\bibinfo {year}
  {2014})}\BibitemShut {NoStop}%
\bibitem [{\citenamefont {Castellini}\ \emph {et~al.}(2018)\citenamefont
  {Castellini}, \citenamefont {Jauslin}, \citenamefont {Rousseaux},
  \citenamefont {Dzsotjan}, \citenamefont {Colas~des Francs}, \citenamefont
  {Messina},\ and\ \citenamefont {Gu{\'e}rin}}]{Castellini2018}%
  \BibitemOpen
  \bibfield  {author} {\bibinfo {author} {\bibfnamefont {A.}~\bibnamefont
  {Castellini}}, \bibinfo {author} {\bibfnamefont {H.~R.}\ \bibnamefont
  {Jauslin}}, \bibinfo {author} {\bibfnamefont {B.}~\bibnamefont {Rousseaux}},
  \bibinfo {author} {\bibfnamefont {D.}~\bibnamefont {Dzsotjan}}, \bibinfo
  {author} {\bibfnamefont {G.}~\bibnamefont {Colas~des Francs}}, \bibinfo
  {author} {\bibfnamefont {A.}~\bibnamefont {Messina}}, \ and\ \bibinfo
  {author} {\bibfnamefont {S.}~\bibnamefont {Gu{\'e}rin}},\ }\href {\doibase
  10.1140/epjd/e2018-90322-5} {\bibfield  {journal} {\bibinfo  {journal} {The
  European Physical Journal D}\ }\textbf {\bibinfo {volume} {72}},\ \bibinfo
  {pages} {223} (\bibinfo {year} {2018})}\BibitemShut {NoStop}%
\bibitem [{\citenamefont {Finkelstein-Shapiro}\ \emph {et~al.}()\citenamefont
  {Finkelstein-Shapiro}, \citenamefont {Mante}, \citenamefont {Sarizosen},
  \citenamefont {Wittenbecher}, \citenamefont {Minda}, \citenamefont {Balci},
  \citenamefont {Pullerits},\ and\ \citenamefont
  {Zigmantas}}]{Finkelstein-Shapiro2020_plex}%
  \BibitemOpen
  \bibfield  {author} {\bibinfo {author} {\bibfnamefont {D.}~\bibnamefont
  {Finkelstein-Shapiro}}, \bibinfo {author} {\bibfnamefont {P.-A.}\
  \bibnamefont {Mante}}, \bibinfo {author} {\bibfnamefont {S.}~\bibnamefont
  {Sarizosen}}, \bibinfo {author} {\bibfnamefont {L.}~\bibnamefont
  {Wittenbecher}}, \bibinfo {author} {\bibfnamefont {I.}~\bibnamefont {Minda}},
  \bibinfo {author} {\bibfnamefont {S.}~\bibnamefont {Balci}}, \bibinfo
  {author} {\bibfnamefont {T.}~\bibnamefont {Pullerits}}, \ and\ \bibinfo
  {author} {\bibfnamefont {D.}~\bibnamefont {Zigmantas}},\ }\href@noop {}
  {\bibinfo  {journal} {arXiv:2002.05642}\ }\BibitemShut {NoStop}%
\bibitem [{\citenamefont {Ribeiro}\ \emph {et~al.}(2018)\citenamefont
  {Ribeiro}, \citenamefont {Mart\'{i}nez-Mart\'{i}nez}, \citenamefont {Du},
  \citenamefont {Campos-Gonzalez-Angulo},\ and\ \citenamefont
  {Yuen-Zhou}}]{Ribeiro2018}%
  \BibitemOpen
\bibfield  {journal} {  }\bibfield  {author} {\bibinfo {author} {\bibfnamefont
  {R.~F.}\ \bibnamefont {Ribeiro}}, \bibinfo {author} {\bibfnamefont {L.~A.}\
  \bibnamefont {Mart\'{i}nez-Mart\'{i}nez}}, \bibinfo {author} {\bibfnamefont
  {M.}~\bibnamefont {Du}}, \bibinfo {author} {\bibfnamefont {J.}~\bibnamefont
  {Campos-Gonzalez-Angulo}}, \ and\ \bibinfo {author} {\bibfnamefont
  {J.}~\bibnamefont {Yuen-Zhou}},\ }\href {\doibase 10.1039/C8SC01043A}
  {\bibfield  {journal} {\bibinfo  {journal} {Chem. Sci.}\ }\textbf {\bibinfo
  {volume} {9}},\ \bibinfo {pages} {6325} (\bibinfo {year} {2018})}\BibitemShut
  {NoStop}%
\bibitem [{\citenamefont {Lindblad}(1976)}]{Lindblad1976}%
  \BibitemOpen
  \bibfield  {author} {\bibinfo {author} {\bibfnamefont {G.}~\bibnamefont
  {Lindblad}},\ }\href@noop {} {\bibfield  {journal} {\bibinfo  {journal}
  {Communications in Mathematical Physics}\ }\textbf {\bibinfo {volume} {48}},\
  \bibinfo {pages} {119} (\bibinfo {year} {1976})}\BibitemShut {NoStop}%
\bibitem [{\citenamefont {Gorini}\ \emph {et~al.}(1976)\citenamefont {Gorini},
  \citenamefont {Kossakowski},\ and\ \citenamefont {Sudarshan}}]{Gorini1976}%
  \BibitemOpen
  \bibfield  {author} {\bibinfo {author} {\bibfnamefont {V.}~\bibnamefont
  {Gorini}}, \bibinfo {author} {\bibfnamefont {A.}~\bibnamefont {Kossakowski}},
  \ and\ \bibinfo {author} {\bibfnamefont {E.~C.~G.}\ \bibnamefont
  {Sudarshan}},\ }\href@noop {} {\bibfield  {journal} {\bibinfo  {journal}
  {Journal of Mathematical Physics}\ }\textbf {\bibinfo {volume} {17}},\
  \bibinfo {pages} {821} (\bibinfo {year} {1976})}\BibitemShut {NoStop}%
\bibitem [{\citenamefont {Knezevic}\ and\ \citenamefont
  {Ferry}(2002)}]{Knezevic2002}%
  \BibitemOpen
  \bibfield  {author} {\bibinfo {author} {\bibfnamefont {I.}~\bibnamefont
  {Knezevic}}\ and\ \bibinfo {author} {\bibfnamefont {D.~K.}\ \bibnamefont
  {Ferry}},\ }\href {\doibase 10.1103/PhysRevE.66.016131} {\bibfield  {journal}
  {\bibinfo  {journal} {Phys. Rev. E}\ }\textbf {\bibinfo {volume} {66}},\
  \bibinfo {pages} {016131} (\bibinfo {year} {2002})}\BibitemShut {NoStop}%
\bibitem [{\citenamefont {Havel}(2003)}]{Havel2003}%
  \BibitemOpen
  \bibfield  {author} {\bibinfo {author} {\bibfnamefont {T.~F.}\ \bibnamefont
  {Havel}},\ }\href {\doibase 10.1063/1.1518555} {\bibfield  {journal}
  {\bibinfo  {journal} {Journal of Mathematical Physics}\ }\textbf {\bibinfo
  {volume} {44}},\ \bibinfo {pages} {534} (\bibinfo {year} {2003})}\BibitemShut
  {NoStop}%
\bibitem [{\citenamefont {Garbe}\ \emph {et~al.}(2020)\citenamefont {Garbe},
  \citenamefont {Bina}, \citenamefont {Keller}, \citenamefont {Paris},\ and\
  \citenamefont {Felicetti}}]{Garbe2020}%
  \BibitemOpen
  \bibfield  {author} {\bibinfo {author} {\bibfnamefont {L.}~\bibnamefont
  {Garbe}}, \bibinfo {author} {\bibfnamefont {M.}~\bibnamefont {Bina}},
  \bibinfo {author} {\bibfnamefont {A.}~\bibnamefont {Keller}}, \bibinfo
  {author} {\bibfnamefont {M.~G.~A.}\ \bibnamefont {Paris}}, \ and\ \bibinfo
  {author} {\bibfnamefont {S.}~\bibnamefont {Felicetti}},\ }\href {\doibase
  10.1103/PhysRevLett.124.120504} {\bibfield  {journal} {\bibinfo  {journal}
  {Phys. Rev. Lett.}\ }\textbf {\bibinfo {volume} {124}},\ \bibinfo {pages}
  {120504} (\bibinfo {year} {2020})}\BibitemShut {NoStop}%
\bibitem [{\citenamefont {Hwang}\ \emph {et~al.}(2015)\citenamefont {Hwang},
  \citenamefont {Puebla},\ and\ \citenamefont {Plenio}}]{Hwang2015}%
  \BibitemOpen
  \bibfield  {author} {\bibinfo {author} {\bibfnamefont {M.-J.}\ \bibnamefont
  {Hwang}}, \bibinfo {author} {\bibfnamefont {R.}~\bibnamefont {Puebla}}, \
  and\ \bibinfo {author} {\bibfnamefont {M.~B.}\ \bibnamefont {Plenio}},\
  }\href {\doibase 10.1103/PhysRevLett.115.180404} {\bibfield  {journal}
  {\bibinfo  {journal} {Physical Review Letters}\ }\textbf {\bibinfo {volume}
  {115}},\ \bibinfo {pages} {180404} (\bibinfo {year} {2015})}\BibitemShut
  {NoStop}%
\bibitem [{\citenamefont {Puebla}\ \emph {et~al.}(2017)\citenamefont {Puebla},
  \citenamefont {Hwang}, \citenamefont {Casanova},\ and\ \citenamefont
  {Plenio}}]{Puebla2017}%
  \BibitemOpen
  \bibfield  {author} {\bibinfo {author} {\bibfnamefont {R.}~\bibnamefont
  {Puebla}}, \bibinfo {author} {\bibfnamefont {M.-J.}\ \bibnamefont {Hwang}},
  \bibinfo {author} {\bibfnamefont {J.}~\bibnamefont {Casanova}}, \ and\
  \bibinfo {author} {\bibfnamefont {M.~B.}\ \bibnamefont {Plenio}},\ }\href
  {\doibase 10.1103/PhysRevLett.118.073001} {\bibfield  {journal} {\bibinfo
  {journal} {Physical Review Letters}\ }\textbf {\bibinfo {volume} {118}},\
  \bibinfo {pages} {073001} (\bibinfo {year} {2017})},\ \bibinfo {note}
  {publisher: American Physical Society}\BibitemShut {NoStop}%
\bibitem [{\citenamefont {Hwang}\ \emph {et~al.}(2018)\citenamefont {Hwang},
  \citenamefont {Rabl},\ and\ \citenamefont {Plenio}}]{Hwang2018}%
  \BibitemOpen
  \bibfield  {author} {\bibinfo {author} {\bibfnamefont {M.-J.}\ \bibnamefont
  {Hwang}}, \bibinfo {author} {\bibfnamefont {P.}~\bibnamefont {Rabl}}, \ and\
  \bibinfo {author} {\bibfnamefont {M.~B.}\ \bibnamefont {Plenio}},\ }\href
  {\doibase 10.1103/PhysRevA.97.013825} {\bibfield  {journal} {\bibinfo
  {journal} {Physical Review A}\ }\textbf {\bibinfo {volume} {97}},\ \bibinfo
  {pages} {013825} (\bibinfo {year} {2018})},\ \bibinfo {note} {publisher:
  American Physical Society}\BibitemShut {NoStop}%
\bibitem [{\citenamefont {Penrose}(1955)}]{Penrose1955}%
  \BibitemOpen
  \bibfield  {author} {\bibinfo {author} {\bibfnamefont {R.}~\bibnamefont
  {Penrose}},\ }\href {\doibase 10.1017/S0305004100030401} {\bibfield
  {journal} {\bibinfo  {journal} {Mathematical Proceedings of the Cambridge
  Philosophical Society}\ }\textbf {\bibinfo {volume} {51}},\ \bibinfo {pages}
  {406} (\bibinfo {year} {1955})}\BibitemShut {NoStop}%
\end{thebibliography}%

\end{document}